\documentstyle[]{book}
\input{psfig.tex}

\let\cchapter=\chapter
\renewcommand{\chapter}{\setcounter{page}{1}\cchapter}

\let\ssection=\section
\renewcommand{\section}{\setcounter{equation}{0}\ssection}

\newcounter{def-number}[chapter]

\newcommand\ie{{\em i.e.,\ }}
\newcommand\nn{\nonumber}

\newcommand{\be}{\begin{enumerate}}
\newcommand{\ee}{\end{enumerate}}
\newcommand{\bi}{\begin{itemize}}
\newcommand{\ei}{\end{itemize}}
\newcommand{\beq}{\begin{equation}}
\newcommand{\eeq}{\end{equation}}
\newcommand{\beqa}{\begin{eqnarray}}
\newcommand{\eeqa}{\end{eqnarray}}

\newcommand{\bra}[1]{\langle{#1}|}
\newcommand{\braket}[2]{\bra{#1}{#2}\rangle}
\newcommand{\eqdef}{\stackrel{\rm def}{=}}
\newcommand{\ket}[1]{|#1\rangle}
\newcommand{\ketbra}[2]{\ket{#1}\bra{#2}}

\newcommand{\ebra}[2]{\langle \exp #1 (#2) |}
\newcommand{\ebraket}[4]{\ebra #1 #2 \exp #3 (#4) \rangle}

\newcommand{\ot}{\otimes}
\newcommand{\op}[1]{\widehat{#1}}
\newcommand{\tr}{{\rm tr}}
\newcommand{\half}{\frac{1}{2}}
\newcommand{\inner}[2]{\langle #1 ,#2\rangle}

\newcommand\mathC{\mkern1mu\raise2.2pt\hbox{$\scriptscriptstyle|$}
		{\mkern-7mu\rm C}}
\newcommand{\mathR}{{\rm I\! R}}                

\renewcommand\[{[\,}                            
\renewcommand\]{\,]}                            

\renewcommand{\H}{{\cal H}}
\newcommand{\K}{{\cal K}}
\renewcommand{\L}{{\cal L}}
\renewcommand{\P}{{\cal P}}
\renewcommand{\S}{{\cal S}}

\newcommand{\U}{{\cal U}}
\newcommand{\UP}{{\cal UP}}
\newcommand{\V}{{\cal V}}
\newcommand{\Vcts}{{\V_{\rm cts}}}

\newcommand{\lbra}{\left(}
\newcommand{\rbra}{\right)}

\newcommand{\hh}[2]{{#1}_{t_1},{#1}_{t_2},\ldots,{#1}_{t_#2}}

\renewcommand{\nn}{\nonumber}
\newcommand{\h}[1]{({#1}_{t_1},{#1}_{t_2},\ldots,{#1}_{t_n})}
\newcommand{\oph}[1]{(\op{#1}_{t_1},\op{#1}_{t_2},\ldots,\op{#1}_{t_n})}
\newcommand{\tp}[1]{\op{#1}_{t_1}\otimes\op{#1}_{t_2}\otimes\ldots\otimes\op{#1}_{t_n} } 
\renewcommand{\P}{\op{P}}
\newcommand{\Q}{\op{Q}}

\renewcommand\H{{\cal H}}

\renewcommand\P{{\cal P}}
\renewcommand\S{{\cal S}}

\renewcommand\half{{\frac 12}}

\newcommand\map{\rightarrow}

\renewcommand\mathR{{\rm I\! R}}

\renewcommand\op[1]{\widehat{#1}}

\renewcommand\tr{{\rm tr}\,}

\newcommand\unit{{\rm 1\kern-3.2pt I}}

\renewcommand\a{\alpha}                         
\renewcommand\b{\beta}                          
\newcommand\g{\gamma}

\renewcommand\l{\lambda}                        
\newcommand\m{\mu}

\newcommand\r{\rho}
\newcommand\s{\sigma}
\newcommand\th{\theta}

\newcommand{\R}{{\rm I\! R}}                

\renewcommand{\S}{{\cal S}}
\renewcommand{\H}{{\cal H}}
\newcommand{\T}{{\cal T}}
\renewcommand{\L}{{\cal L}}
\renewcommand{\P}{\op{P}}

\renewcommand{\bra}[1]{\langle{#1}|}
\renewcommand{\ket}[1]{|#1\rangle}
\renewcommand{\braket}[2]{\bra{#1}{#2}\rangle} 
\renewcommand{\ketbra}[2]{\ket{#1}\bra{#2}}

\begin{document}

\begin{titlepage}

\begin{center}
{\large\bf Continuous Time and Consistent Histories\\[6pt]
        }
\end{center}

\vspace{0.8 truecm}
\begin{center}
        Konstantina  Savvidou\\[30pt]
        Theoretical Physics Group\\
        The Blackett Laboratory\\
        Imperial College of Science, Technology \& Medicine\\
        South Kensington\\
        London SW7 2BZ\\
\end{center}

\end{titlepage}

\tableofcontents

\chapter*{Acknowledgements}
I would like to greatfully thank  my supervisor Chris Isham for giving me the chance to work in this area of research and for teaching me how to realise ideas into concrete work. I would also like to thank my family and my friends Giannis Kouletsis, Kostas Viglas and Andreas Zoupas. I am especially greatful to Charis Anastopoulos for the encouragement and help through these years. I would like to thank Nikos Plakitsis for supporting me always. I gratefully acknowledge support from the NATO Research Fellowships department of the Greek Ministry of Economy.

\chapter*{Abstract}

We discuss the case of histories labelled by a continuous time parameter in the {\em History Projection Operator} consistent-histories quantum theory. In this approach---an extention to the generalised consistent histories theory---propositions about the history of the system are represented by projection operators on a Hilbert space. A continuous time parameter leads to a history algebra that is isomorphic to the
canonical algebra of a quantum {\em field\/} theory. We describe how the appropriate representation of the history algebra may be
chosen by requiring the existence of projection operators that
represent propositions about time averages of the energy. We define the action operator for the consistent histories
formalism, as the quantum analogue of the classical action
functional, for the simple harmonic oscillator case. We
show that the action operator is the generator of two types of time
transformations that may be related to the two laws of time-evolution of the standard quantum theory: the
state-vector reduction and the unitary time-evolution. We
construct the corresponding classical histories and
demonstrate the relevance with the quantum histories; we demonstrate how the requirement of the temporal logic structure of the theory is sufficient for the definition of classical histories.
Furthermore, we show the relation of the action operator to the
decoherence functional which describes the dynamics of the system. Finally, the discussion is extended to give a
preliminary account of quantum field theory in this approach to the
consistent histories formalism.

\chapter{Introduction}
In classical Newtonian theory, time is introduced as an
external parameter; and in all the existing approaches to
quantum theory, the treatment of time is inherited from the
classical theory. On the other hand, general relativity
treats time as an internal parameter of the theory: in
particular, it is one of the coordinates of the spacetime
manifold. When we combine the two theories in quantum
gravity, this essential difference in the treatment of time
appears as a major problem---one of the aspects of what is
known as the `Problem of Time'. One of the directions
towards a solution of the problem is to construct
`timeless' quantum theories, {\em i.e.} theories where time
is not a fundamental ingredient of the theory.
\vspace*{20pt}

One such formalism is the consistent histories approach to
quantum theory in which time appears as the label on a
time-ordered sequence of projection operators which
represents a `history' of the system. In the original
scheme by Gell-Mann and Hartle \cite{Gri84,Omn88a,GH90b},
the crucial object is the decoherence function written as
$d(\alpha,\beta)= {\rm tr}(\tilde C_\alpha^\dagger\rho
\tilde C_\beta)$ where $\rho$ is the initial
density-matrix, and where the {\em class operator\/}
$\tilde C_\alpha$ is defined in terms of the standard
Schr\"odinger-picture projection operators $\alpha_{t_i}$
as
\begin{equation}
 \tilde C_\alpha:=U(t_0,t_1)\alpha_{t_1} U(t_1,t_2)
    \alpha_{t_2}\ldots U(t_{n-1},t_n)\alpha_{t_n}U(t_n,t_0)
\end{equation}
where $U(t,t')=e^{-i(t-t')H/\hbar}$ is the unitary
time-evolution operator from time $t$ to $t'$. Each
projection operator $\alpha_{t_i}$ represents a proposition
about the system at time $t_i$, and the class operator
$\tilde C_\alpha$ represents the composite history
proposition ``$\alpha_{t_1}$ is true at time $t_1$, and
then $\alpha_{t_2}$ is true at time $t_2$, and then \ldots,
and then $\alpha_{t_n}$ is true at time $t_n$''.
\vspace*{20pt}

The motivation for the work that will be presented here,
may be elucidated in several {\em key points\/} about the
consistent histories theory construction.

1. The consistent histories approach allows the description
of an approximately classical domain emerging from the
macroscopic behaviour of a closed physical system, as well
as its microscopic properties in terms of the conventional
Copenghagen quantum mechanics. This is possible through the
decoherence condition: the requirement for `decoherence'
(negligible interference between histories leads to the
assignment of a probability measure) selects a consistent
set of histories that can be represented on a classical
(Boolean) logic lattice, thus having a classical logical
structure. Hence in the consistent histories theory,
emphasis must be given to the observation that, although in
atomic scales a system is described by quantum mechanics,
it may also be described by classical mechanics and
ordinary logic. Therefore a more refined logical structure
seems to be a necessary part of any consistent histories
formalism. However, the Gell-Mann and Hartle approach lacks
the logical structure of standard quantum mechanics in the
sense that the fundamental entity ({\em i.e.}history) for
the description of the system is not represented by a
projector in the standard Hilbert space representation: as
a product of (generically, non-commuting) projection
operators, the class-operator $\tilde C_\alpha$,
representing a history, is not itself a projector.

2. This difference between the representation of
propositions in standard quantum mechanics and in the
history theory was resolved in the alternative approach of
the `History Projection Operator' (or HPO for short) theory
\cite{Isham94,IL94}, in which the history proposition
``$\alpha_{t_1}$ is true at time $t_1$, and then
$\alpha_{t_2}$ is true at time $t_2$, and then \ldots, and
then $\alpha_{t_n}$ is true at time $t_n$'' is represented
by the {\em tensor product\/}
$\alpha_{t_1}\otimes\alpha_{t_2}\otimes
\cdots\otimes\alpha_{t_n}$ which, unlike $\tilde C_\alpha$,
{\em is\/} a genuine projection operator on the tensor
product of copies of the standard Hilbert space ${\cal V}_n
= {\cal H}_{t_1}\otimes{\cal
H}_{t_2}\otimes\cdots\otimes{\cal H}_{t_n}$. Hence the
`History Projection Operator' formalism restores the
quantum logic structure as it is in the case of single-time
quantum theory.

3. However, the introduction of the tensor product ${\cal
H}_{t_1}\otimes{\cal H}_{t_2}\otimes\cdots\otimes{\cal
H}_{t_n}$ led to a quantum theory where the notion of time
appears mainly via its partial ordering property
(quasi-temporal behaviour). In particular, we do not have a
clear notion of time evolution in the sense that there is
no natural way to express the time translations from one
time slot---that refers to one copy of the Hilbert space
${\cal H}_t$---to another one, that refers to another copy
${\cal H}_{t^{\prime}}$. As we shall see, the situation
changes when the {\em the continuous limit\/} of such
tensor products is introduced: henceforward, time appears
uniformly in a continuous way.

4. One of the original problems in the development of the
HPO theory was the lack of a clear physical meaning of the
quantities involved. The introduction of the history group
by Isham and Linden \cite{IL95} made a significant step in
this direction in the sense that, the spectral projectors
of the history Lie algebra represent propositions about
phase space observables of the system. Furthermore, it
transpired that the history algebra for one-dimensional
quantum mechanics is {\em infinite\/} dimensional---in
fact, it is isomorphic to the canonical commutation algebra
of a standard quantum {\em field\/} theory in one spatial
dimension. This suggested that it would be profitable to
study the history theory using tools that are normally
employed in quantum field theory. We will use such tools
extensively in what follows.

5. The choice of the continuous-time treatment introduced
in the definition of the history algebra (history
commutation relations at {\em unequal times\/}) by a
delta-function, has a striking consequence: the physical
observables of the theory are intrinsically time-averaged
quantities; this means that the physical quantities cannot
be defined at {\em sharp moments in time\/}. This is an
important feature of the HPO theory. In this respect, it is
closer to quantum field theory formalisms but with the
essential difference that the time (spacetime) smearing
does not appear only as a mathematical requirement but is
also an intrinsic property of the fundamental elements of
the theory.

This latter result, together with the preceding reasoning,
was the starting point for the work that will be presented
here. As we shall see, the introduction of the
continuous-time treatment enables the definition of time
transformations in the HPO theory, leading finally the
notion of a {\em time flow\/}.

We will now give a brief description of the contents of
this work.

In chapter 2, we summarise the generalised consistent
histories theory in the form originally developed by
Gell-Mann and Hartle. We also show how a history is
represented in standard quantum mechanics, using the
underlying logical structure of the theory. We then give a
detailed presentation of the History Projection Operator
theory based on the ideas of Isham \cite{Isham94}. In
particular, we emphasise the logical structure of the
theory, which is one of the key descriminating factors from
previous consistent histories formalisms.

In chapter 3, we explain the choice of treating time as a
continuous parameter. The construction of the history group
is an important part of the HPO theory, therefore we
present its definition and the original attempt to find the
representation space of the history algebra for the example
of a particle moving on a line $\mathcal R$, as presented
in \cite{IL95}. We comment on the observation that the
history commutation relations are identical to the ones for
the one spatial dimention quantum field theory. We then
embark on a more physicaly motivated construction, based on
the fact that the requirement for the existence of a
Hamiltonian operator properly defined on the history space
uniquely selects the history algebra representation space.
In particular we examine the example of the simple harmonic
oscillator in one dimention. As we explained previously, in
HPO the interesting question that arises is how the
Schr\"odinger-picture objects with different time
labels---refering to the corresponding copies of the
standard Hilbert space---are related. This work is the result of the collaboration between the authors of the article \cite{1}. The work presented in the following chapters is published in \cite{S98}. 

In chapter 4 we will
show that there is a transformation law `from one Hilbert
space to another'. The generator of these time
transformations is the `action operator', a quantum
analogue of the classical action functional. The definition
of Heisenberg-picture operators will be used to demonstrate
the time transformation law.

The main theme of chapter 5 is exactly the time
transformation structure of HPO theory. In particular, we
will show that there exist two types of time
transformation, generated by the kinematical (Liouville
operator) and the dynamical (Hamiltonian operator) part of
the action operator. We will try to interpret the two-fold
time law comparing it with the two types of time evolution
of standard quantum theory, {\em i.e.}, the state-vector
reduction and the unitary time evolution.

It is interesting to examine how this novel structure with
respect to the two time transformations in HPO appear in
classical case. To this end, we define classical histories
as an analogue of the quantum ones. Furthermore, we will
show that, taking into account the temporal logic structure
of the theory, classical histories can be defined without
any reference to the quantum case.

The dynamics of the theory is described by the decoherence
functional: it is natural to seek then the appearance of
the action operator in its expression. We first present a
summary of the use of coherent states for the definition of
the decoherence functional as was originally presented in
\cite{IL95}. Then we show that the operators involved have
a functional relation with the action operator.

In chapter 6, we extend the discussion to the HPO theory of
a free scalar field. In particular, starting from the
quantum mechanics history group, we write a possible
candidate for the quantum field theory history algebra. The
question of external Lorentz invariance is examined. We
comment on the intriguing result that in HPO two possible
Poincar\'e groups appear, as a consequence of the two types
of time transformation.




%



\chapter{A Review of the Consistent Histories}

\section{Introduction}

The origins of the consistent histories theory
lies in the attempt, introduced by Everett, to
apply quantum mechanics to closed systems. The
usual Copenghagen formulation of quantum mechanics
is inadequate for quantum cosmology as it assumes
a division of the universe into `observer' and
`observed', and for the early universe it posits
an external `classical domain'.

The post-Everett formulation of quantum mechanics
stresses the consistency of probability sum rules
as the primary criterion for determining which
sets of histories may be assigned probabilities,
and the {\em decoherence} (absence of interference
between individual histories) as a sufficient
condition for the consistency of probability sum
rules. Such sets of histories are called
`consistent' or `decoherent' and can be
manipulated according to the rules of ordinary
Boolean logic.

The consistent histories theory introduces a new
treatment of the notion of time that opens up the
possibility of eventually finding a different way
to address the relevant problems in quantum
gravity. In this sense, the Hamiltonian quantum
mechanics is constructed by choosing one set of
spacelike surfaces to define time. Hence it is
restricted to a particular choice for the
direction of time. Hartle partly resolved this
problem by using the sum-over-histories
formulation of quantum mechanics to bring
histories in a spacetime form so that the theory
does not require a priviledged notion of time. In
addition it works for a large variety of temporal
coarse- grainings such as spacetime regions.
However, it is restricted only to configuration
space histories.

Isham suggested a refinement of the Gell-Mann and
Hartle axioms for a generalized histories approach
by constructing analogues of the lattice structure
employed in standard quantum logic. Its
quasitemporal structure is coded in a partial
semigroup of temporal supports incorporated in the
lattice of history propositions by the
correspondance of a temporal support to each
history proposition. This treatment of the notion
of time is of great significance for quantum
gravity and quantum field theories in curved
spacetime.

\section{The Formalism of Consistent Histories}

The usual `Copenhagen' formulation of quantum
mechanics is inadequate for the description of
possible histories for the universe because of the
absence in this case of the external observer.
There thereby arises a need to be able to assign
probabilities about alternative histories of a
subsystem without using the notion of a
measurement procedure as a necessary ingredient.
The idea that was developed in this direction was
that the primary criterion for the assignment of
probabilities is the consistency of probability
sum rules, and the sufficient condition for this
is the absence of quantum mechanical interference
between individual histories, {\em i.e.}, the
notion of decoherence.

After the original attempt for a quantum
description of the universe by Everett, there
followed the construction of a generalized quantum
mechanics that led to the formulation of the
history theory approach.

\subsection{Histories in Standard Quantum Theory}
It is useful to summarise very briefly how
`histories' are understood in the conventional
interpretation of an open, Hamiltonian quantum
system that is subject to measurements by an
external observer.

To this end, let $U(t_1,t_0)$ denote the unitary
time-evolution operator from time $t_0$ to $t_1$;
\ie $U(t_1,t_0)=e^{-i(t_1-t_0)H/\hbar}$. Then, in
the Schr\"odinger picture, the density operator
state $\r(t_0)$ at time $t_0$ evolves in time
$t_1-t_0$ to $\r(t_1)$, where
 \beq
\r(t_1)=U(t_1,t_0)\r(t_0)U(t_1,t_0)^{\dag}=U(t_1,t_0)\r(t_0)U(t_1,t_0)^{-1}.
\eeq Suppose that a measurement is made at time
$t_1$ of a property represented by a projection
operator $P$. Then the probability that the
property will be found is \beq {\rm
Prob}\big(P=1;\rho(t_1)\big)=\tr\big(P\r(t_1)\big)
=
    \tr\big(PU(t_1,t_0)\r(t_0)U(t_1,t_0)^{\dag}\big) =
        \tr\big(P(t_1)\r(t_0)\big)
\eeq where \beq
    P(t_1):=U(t_1,t_0)^{\dag}P(t_0)U(t_1,t_0)
\eeq is the Heisenberg-picture operator defined
with respect to the fiducial time $t_0$. If the
result of this measurement is kept then, according
to the Von Neumann `reduction' postulate, the
appropriate density matrix to use for any further
calculations is \beq \r_{\rm
red}(t_1):={P(t_1)\r(t_0)P(t_1)\over\tr\big(P(t_1)\r(t_0)\big)}.
\eeq

    Now suppose a measurement is performed of a second observable $Q$
at time $t_2>t_1$. Then, according to the above,
the {\em conditional\/} probability of getting
$Q=1$ at time $t_2$ given that $P=1$ was found at
time $t_1$ (and that the original state was
$\r(t_0)$) is \beq
    {\rm Prob}\big(Q=1|P=1{\rm\ at\ }t_1;\r(t_0)\big)=
        \tr\big(Q(t_2)\r_{\rm red}(t_1)\big)=
    {\tr\big(Q(t_2)P(t_1)\r(t_0)P(t_1)\big)\over\tr\big(P(t_1)\r(t_0)\big)}.
\eeq The probability of getting $P=1$ at $t_1$
{\em and\/} $Q=1$ at $t_2$ is this conditional
probability multiplied by ${\rm
Prob}\big(P=1;\r(t_1)\big)$, \ie \beq {\rm
Prob}\big(P=1{\rm\ at\ }t_1{\rm {\em\ and\
}}Q=1{\rm\ at\ }t_2;
    \r(t_0)\big)=\tr\big(Q(t_2)P(t_1)\r(t_0)P(t_1)\big).
\eeq Generalising to a sequence of measurements of
propositions $\a_{t_1},\a_{t_2},\ldots,\a_{t_n}$
at times $t_1,t_2,\ldots,t_n$, the joint
probability of finding all the associated
properties is \beqa \lefteqn{{\rm
Prob}\big(\a_{t_1}=1{\rm\ at\ }t_1 {\rm{\em\ and}\
}
    \a_{t_2}=1{\rm\ at\ }t_2{\rm{\em\ and\ldots}}
    \a_{t_n}=1{\rm\ at\ }t_n;\r(t_0)\big)=}\ \ \ \ \ \ \      \nonumber\\
    &&\ \ \tr\big(\a_{t_n}(t_n)\ldots\a_{t_1}(t_1)\r(t_0)
        \a_{t_1}(t_1)\ldots\a_{t_n}(t_n)\big)
                                                    \label{Prob:a1_an}
\eeqa The conditions that must be satisfied for
the probability assignments Eq.\
(\ref{Prob:a1_an}) to be consistent are presented
below, in the context of the Gell-Mann and Hartle
axioms.

\subsection{The Gell-Mann and Hartle Generalised Consistent Histories Approach}
 The generalised consistent-histories approach to quantum theory can be
formulated in several different ways. In the
original scheme by Gell-Mann and Hartle
\cite{Gri84,Omn88a,GH90b}, the main assumption of
the consistent-histories interpretation of quantum
theory is that, under appropriate conditions, a
probability assignment is still meaningful for a
{\em closed\/} system, with no external observers
or associated measurement-induced state-vector
reductions (thus signalling a move from
`observables' to `beables'). The satisfaction or
otherwise of these conditions is determined by the
behaviour of the {\em decoherence functional\/}
$d_{\r,H}(\a,\b)$ which, for the pair of sequences
of projection operators $\a:=\h{\a}$ and
$\b:=\h{\b}$ is defined as
\begin{equation}
    d_{\r,H}(\alpha,\beta)=
{\rm tr}(\tilde C_\alpha^\dagger\rho \tilde
C_\beta) \label{Def:d}
\end{equation}
where $\rho$ is the initial density-matrix, $H$ is
the Hamiltonian, and where the {\em class
operator\/} $\tilde C_\alpha$ is defined in terms
of the standard Schr\"odinger-picture projection
operators $\alpha_{t_i}$ as
\begin{equation}
    \tilde C_\alpha:=U(t_0,t_1)\alpha_{t_1} U(t_1,t_2)
    \alpha_{t_2}\ldots U(t_{n-1},t_n)\alpha_{t_n}U(t_n,t_0),
                                                    \label{Def:C_a}
\end{equation}
where $U(t,t')=e^{-i(t-t')H/\hbar}$ is the unitary
time-evolution operator from time $t$ to $t'$.
Each projection operator $\alpha_{t_i}$ represents
a proposition about the system at time $t_i$, and
the class operator $\tilde C_\alpha$ represents
the composite history proposition ``$\alpha_{t_1}$
is true at time $t_1$, and then $\alpha_{t_2}$ is
true at time $t_2$, and then \ldots, and then
$\alpha_{t_n}$ is true at time $t_n$''.

At this point it is useful to gather together a
few definitions that can be conveniently
associated with these ideas.
\bi
\item  A {\em homogeneous history\/}
is any time-ordered sequence $\oph{\a}$ of
projection operators.

\item  A homogeneous history  $\b:=\oph{\b}$ is {\em coarser\/} than
another history $\a:=\oph{\a}$ if, for every
$t_i$, $\op\a_{t_i}\leq\op\b_{t_i}$ where $\leq$
denotes the usual ordering operation on the space
of projection operators, \ie $\P\leq\Q$ means that
the range of $\P$ is a subspace of the range of
$\Q$ (this includes the possibility that $\P=\Q$
so that, in particular, every homogeneous history
is trivially coarser than itself). This relation
on the set of homogeneous histories is a partial
ordering \footnote{A relation $\leq$ on a set $X$
is a {\em partial ordering\/} if it satisfies the
conditions (i) for all $x\in X$, $x\leq x$; (ii)
$x\leq y$ and $y\leq x$ implies $x=y$; and (iii)
$x\leq y$ and $y\leq z$ implies $x\leq z$.}.

\item Two homogeneous histories  $\a:=\oph{\a}$ and $\b:=\oph{\b}$ are
{\em disjoint\/} if, for at least one time point
$t_i$, $\op\b_{t_i}$ is disjoint from
$\op\a_{t_i}$, \ie the ranges of these two
projection operators are orthogonal subspaces of
$\H$.

\item In calculating a decoherence functional it may be necessary to
go outside the class of homogeneous histories to
include {\em inhomogeneous\/} histories. A history
of this type arises as a logical `or' (denoted
$\vee$) operation on a pair of disjoint
homogeneous histories $\a:=\oph{\a}$ and
$\b:=\oph{\b}$. Such a history $\a\vee\b$ is
generally {\em not\/} itself a collection of
projection operators (\ie it is not homogeneous)
but, when computing the decoherence functional, it
is represented by the operator
$\op{C}_{\a\vee\b}:=\op{C}_\a+\op{C}_\b$. The
coarse-graining relations $\a\leq\a\vee\b$ and
$\b\leq\a\vee\b$ are deemed to apply to this
disjoint `or' operation. The `negation' operation
$\neg$ also usually turns a homogeneous history
into an inhomogeneous history, with
$\op{C}_{\neg\a}:=\op{1}-\op{C}_\a$. \ei A brief
description of the elements of the theory follows.

\subsection{The Gell-Mann and Hartle axioms}

The Gell-Mann and Hartle axioms \cite{Har93a}
postulate a new approach to quantum theory in
which the notion of history has a fundamental
role; \ie a `history' in this generalised sense
can be an irreducible entity in its own right, not
necessarily derived from time-ordered strings of
single-time propositions. These axioms and
definitions are essentially as follows:
\be
\item The fundamental ingredients in the theory are a space of {\em
histories\/} and a space of {\em decoherence
functionals\/} which are complex-valued functions
of pairs of histories. The value $d(\a,\b)$ of
such a decoherence functional $d$ is a measure of
the extent to which the histories $\a$ and $\b$
`interfere' with each other.

\item The set of histories possesses a partial order $\leq\,$. If
$\a\leq\b$ then $\b$ is said to be {\em coarser\/}
than $\a$, or a {\em coarse-graining\/} of $\a$;
dually, $\a$ is a {\em finer\/} than $\b$, or a
{\em fine-graining\/} of $\b$ . Heuristically this
means that $\a$ provides a more precise
specification than $\b$.

\item A history $\a$ is defined
to be {\em fine-grained\/} if the only histories
$\b$ for which $\b\leq\a$ are $0$ or $\a$ itself.
In standard quantum theory, histories of this type
arise as time-ordered sequences of projection
operators whose ranges are all one-dimensional
subspaces of the Hilbert space. In general, the
fine-grained histories are the sets of exhaustive,
alternative histories of a closed system which are
the most refined description to which one can
contemplate assigning probabilities.

\item The $\sl{allowed}$ $\sl{ coarse}$ $\sl{ grainings}$. The operation of coarse
graining partitions a set of fine-grained
histories into an exhaustive set of exclusive
classes $\{c_{\alpha }\}$, its class being a
coarse-grained history.

\item There is a notion of two histories $\a,\b$ being {\em
disjoint\/}, written $\a\perp\b$. Heuristically,
if $\a\perp\b$ then if either $\a$ or $\b$ is
`realised' the other is automatically excluded.

\item There is a {\em unit\/} history $1$ (heuristically, the history
that is always realised) and a {\em null
history\/} $0$ (heuristically, the history that is
never realised). For all histories $\a$ we have
$0\leq\a\leq 1$.

\item Two histories $\a,\b$ that are disjoint can be
combined to form a new history $\a\vee\b$
(heuristically, the history `$\a$ {\em or\/}
$\b$') which is the least upper bound of $\a$ and
$\b$ with respect to the partial ordering $\leq$.

\item A set of  histories $\a^1,\a^2,\ldots,\a^N$ is said to be
{\em exclusive\/} if $\a^i\perp\a^j$ for all
$i,j=1,2,\ldots,N$. The set is {\em exhaustive\/}
(or {\em complete\/}) if it is exclusive and if
$\a^1\vee\a^2\vee\ldots\vee\a^N=1$.

\item The $\sl{decoherence}$ $\sl{ functional}$ measures interference between the
members of a coarse-grained set of histories. The
decoherence functional is a complex-valued
functional, $d(\alpha^{'},\alpha)$, defined for
each pair of histories in a coarse-grained set
$\{\alpha \}$. It must satisfy the following
conditions:

  i) Hermiticity: $d(\alpha^{'},\alpha)= d^*(\alpha^{'},\alpha)$

  ii) Positivity: $d(\alpha,\alpha)\geq 0$

  iii) Normalization: $\sum_{\alpha^{'},\alpha} d(\alpha^{'},\alpha)=1$

\ee \noindent

    It is important to note that this axiomatic scheme is given a
physical interpretation only in relation to {\em
consistent\/} sets of histories. A complete set
$\cal C$ of histories is said to be (strongly)
consistent with respect to a particular
decoherence functional $d$ if $d(\a,\b)=0$ for all
$\a,\b\in{\cal C}$ such that $\a\ne\b$. Under
these circumstances, $d(\a,\a)$ is given the
physical interpretation as the {\em probability\/}
that the history $\a$ will be `realised'. The
Gell-Mann and Hartle axioms then guarantee that
the usual Kolmogoroff probability sum rules will
be satisfied.

When we consider Hamiltonian quantum mechanics the
set of histories are represented by chains of
projections onto exhaustive sets of orthogonal
subspaces of a Hilbert space. Then the
fine-grained histories correspond to the possible
sequences of sets of projections onto a complete
set of states, a set at every time. The set of
coarse-grained histories consist of sequences of
independent alternatives at definite moments of
time so that every history can be represented as a
chain of projections.

In order to develop history theory to the fully
4-dimensional spacetime form that does not need a
priviledged notion of time, Hartle generalised
quantum mechanics by enlarging the set of
alternatives to include spacetime ones, not
necessarily defined on spacelike surfaces. He then
incorporated the dynamics of the theory by using
the spacetime path integrals in the decoherence
functional, thus succeeding in a construction that
allows plurality in the selection of different
temporal coarse grainings.

The sum over histories approach is described with
the familiar three elements of a history theory.
The fine-grained histories are paths in a
configuration space of generalized coordinates
$\{q^{i}\}$, expressed as single-valued functions
of the physical time. The operation of
coarse-graining can be made with an especially
natural partition of the configuration space into
a mutually exclusive and exhaustive set of subsets
at each moment of time. The decoherence functional
can be written in a path-integral form as
\begin{equation}
d(\alpha^{'},\alpha)=\int_{C_{\alpha'}}\delta
q'\int_{C_{\alpha}}\delta q\delta (q_{f'}-q_{f})
e^{i/h\left(S\[q'(\tau)\]- S\[q(\tau)\]\right)}
\rho (q'_{0},q_{0})
\end {equation}
for an interval of time t=0 to t=T. The integrals
are defined over paths that begin at a point
$q_{0}$ at t=0, end at a point $q_{f}$ at t=T and
lie in the class $c_{\alpha}$, and $\rho
(q'_{0},q_{0})$ is a density matrix.The
integration with the primes is defined in analogy.
For the case of the above partition, coming with a
fixed choice of time, the decoherence functional
coincides with that of the Hamiltonian quantum
mechanics histories.

In heuristic sense, the sum-over-histories
approach provides a purely covariant framework for
the treatment of field theories because the path
integration is over fields defined over
spacetime.The use of the path integral method in
this case has the ability to accomodate various
choices of temporal structures as encoded within
the different spacetime coarse grainings. On the
other hand, the theory is restricted only for
configuration space histories. This means that
fewer sets of coarse-grained histories are
possible since there is a unique set of fine
grained histories.

\section{The History Projection Operator Approach: the Discrete-Time Case}
As was illustrated in the previous paragraph, the
consistent histories programme affords the
possibility of escaping the measurement problem
and the concept of state-vector reduction induced
by an external observer associated with the
Copenghagen interpetation of quantum theory. In
addition, the fact that the notion of history can
be used as a fundamental theoretical entity rather
than a time-ordered string of events, enables a
novel way of addressing the problem of time in
quantum gravity situations. The `quasi-temporal'
nature of the consistent histories formalism
inspired Isham to further develop a consistent
histories approach in such a way that the theory
is equipped with a generalisation of the quantum
logic structure of the standard quantum theory.

Starting from the ideas of Mittlestaedt and
Stachow on the logic of sequential propositions,
Isham's key idea was the observation that the
statement that a certain universe ({\em i.e.},
history) is `realised' is itself a proposition,
and therefore the set of all such histories might
possess a lattice structure analogous to the
lattice of single-time propositions in standard
quantum logic. In particular, a (general) history
proposition should be represented by a projection
operator in some Hilbert space. In the Gell-Mann
and Hartle approach this is exactly {\em not\/}
the case since the C-representation of a history
defined as $\op{C}_\a:=$
$\op\a_{t_n}(t_n)\op\a_{t_{n-1}}(t_{n-1})\ldots\op\a_{t_1}(t_1)$,
the product of (Heisenberg picture) projection
operators $\op\a_{t_k}(t_k)$ is usually not itself
a projection operator.

\subsection{A History Version of Standard Quantum Theory}
 One of the main aims was to find candidates for the
`history analogues' of the lattice $\L$ and the
state-space $\cal R$ of the standard Hamiltonian
theory. These analogues of $\L$ and $\cal R$ is a
space $\UP$ of history-propositions and a space
$\cal D$ of decoherence functionals. The
construction of such a general scheme was
motivated by the special example of a history
version of standard quantum logic. By this is
meant a generalisation of the ideas in paragraph
2.2 in which strings of projection operators are
replaced by strings of single-time propositions
belonging to the lattice $\L$ of some `standard'
quantum logic theory.

In the quantum-logic version of the history theory
we consider a system with a lattice $\L$ of
single-time propositions, and define a {\em
history filter\/} to be any finite collection
$\h{\a}$ of single-time propositions
$\a_{t_i}\in\L$ which is time-ordered in the sense
that $t_1<t_2<\ldots<t_n$. Thus, in the special
case where $\L$ is identified with the lattice
$P(\H)$ of projection operators on a Hilbert space
$\H$, a history filter is what we called a
homogeneous history above.

    In the case of standard quantum logic, a history filter is a
time-labelled version of what Mittelstaedt and
Stachow call a {\em sequential
conjunction\/}\cite{Mit77,Sta80,Sta81} \ie it
corresponds to the proposition `$\a_{t_1}$ is true
at time $t_1$, {\em and then\/} $\a_{t_2}$ is true
at time $t_2$, {\em and then\/} $\ldots$ {\em and
then\/} $\a_{t_n}$ is true at time $t_n$'. The
phrase `history filter' is intended to capture the
idea that each single-time proposition $\a_{t_i}$
in the collection $\h{\a}$ serves to `filter out'
the properties of the system that are realised in
this potential history of the universe.

    It is important to be able to manipulate history filters that are
associated with {\em different\/} sets of time
points. To this end, it is useful to think of a
history filter as something that is defined at
{\em every\/} time point but which is `active'
only at a finite subset of points. This can be
realised mathematically by defining it to be equal
to the trivial proposition at all but the active
points. More precisely, in standard quantum logic
we shall define a history filter $\a$ to be an
element of the space ${\cal F}(\T,\L)$ of maps
from the space of time points $\T$ (in the present
case, the real line $\R$) to the lattice $\L$ with
the property that each map is (i) equal nowhere to
the null single-time proposition, and (ii) equal
to the unit single-time proposition for all but a
finite set of $t$ values. It will be convenient to
append to this space the null history filter which
is defined to be the null single-time proposition
at all points $t\in\T$.

    It follows that, in a standard quantum theory realised on a
Hilbert space $\H$, a history filter (\ie a
homogeneous history) is represented by an element
$\a$ of the space of functions ${\cal
F}(\T,P(\H))$ where $\op\a_t$ (the value of the
map $\a$ at $t\in\T$) is equal to the unit
operator for all but a finite set of time points
$t\in\T$. Objects of this type can be regarded as
projection operators on the weak direct sum ${\cal
F}(\T,\H)$ of $\H$-valued functions on $\T$.

In the context of quantum cosmology, a history
filter is a possible `{\em universe\/}' complete
with whatever quasi-temporal attributes it may, or
may not, possess. For this reason, the set of all
history filters in the general theory will be
denoted $\U$; in the case of standard quantum
logic we will write $\U(\L):={\cal F}(\T,\L)$ to
indicate the underlying lattice $\L$ of
single-time propositions.

 The temporal properties of a history filter
$\a\in\U(\L)$ are encoded in the finite set of
time points at which it is active; \ie the points
$t\in\T$ such that $\a_t\ne 1$. This motivates the
following definitions:
\be

\item The set of $t\in\T$ for which $\a_t\ne1$ is called the {\em
      temporal support\/}, or just {\em support\/}, of $\a\in\U(\L)$,
      and is denoted $\s(\a)$.

\item The set of all possible temporal supports will be denoted $\S$;
      in the present case this is just the set of all ordered finite
      subsets of $\T=\R$.

\item The support of the null history is defined to be the empty
      subset of $\R$.
\ee The fact that the essential temporal
properties of the space of history filters
$\U(\L)$ in standard Hamiltonian quantum theory is
reflected in its set of temporal supports raised
the possibility to construct a theory that is more
general than the standard theory and in which the
quasi-temporal structure is reflected in the
structurre of the support space. Furthermore, in
the general case in order for the `or' and
`not'operations to be included, $\U$ is extended
to a larger space $\UP$ of all history
propositions which has the structure of an
orthocomplemented lattice.

\subsection{The HPO-theory for Standard Quantum Theory}
The next step was to construct an operator
representation of standard quantum theory in which
every history proposition {\em is\/} represented
by a genuine projection operator, thus the whole
$\UP$ can be identified with the projection
lattice of some new Hilbert space. To accomplice
this Isham examined the special case where
$\L=P(\H)$, {\em i.e.}, a Hilbert space based
quantum system rather than a general lattice $\L$.
Still, for a general history theory called an HPO
theory (History Projection Operator theory), the
space $\UP$ is an ortho-complemented lattice that
can be represented by projection operators.

As we have explained previously, $\op{C}_\a$ is
not a projection operator, and therefore it is not
part of the propositional lattice associated with
the Hilbert space $\H$ on which it is defined.
This makes it difficult to know what is the
$C$-representative of, for example, $\a\vee\b$
when $\a$ and $\b$ are {\em not\/} disjoint.
Indeed, if $\P$ and $\Q$ are projection operators,
the product $\P\Q$ generally fails to be so.
However, the {\em tensor\/} product $\P\otimes\Q$
{\em is\/} a projection operator, and is hence a
candidate to represent the two-time homogeneous
history $(\P,\Q)$.
 More generally, if we consider the set
$\U_{\{t_1,\ldots,t_n\}}$ of all homogeneous
histories with (for the moment) a fixed support
$\{t_1,t_2,\ldots,t_n\}$, we can represent any
such $\a=\oph{\a}$ with the tensor product \beq
    \th\oph{\a}:=\tp{\a}                          \label{Def:th}
\eeq which acts on the tensor-product space
$\otimes_{t\in\{t_1,\ldots, t_n\}}\H_t$ of $n$
copies of $\H$.

    That the tensor product appears in a natural way can be seen from
the following observation. In constructing the
decoherence functional, the map \beq
    \oph{\a}\mapsto \tr\big(\op\a_{t_n}(t_n)\op\a_{t_{n-1}}(t_{n-1})
        \ldots\op\a_{t_1}(t_1)\op{B}\big)
\eeq is {\em multilinear\/} with respect to the
vector space structure of
$\oplus_{t\in\{t_1,\ldots t_n\}}B(\H)_t$ for any
$\op{B}\in B(\H)$. However, the fundamental
property of the tensor product of a finite
collection of vector spaces $V_1,V_2,\ldots, V_n$
is that any multilinear map $\mu:V_1\times
V_2\times\ldots\times V_n\map W$ to a vector space
$W$ factorises uniquely through the tensor product
to give the chain of maps \beq
    V_1\times V_2\times\ldots\times V_n\buildrel\th\over\rightarrow
        V_1\otimes V_2\otimes\ldots\otimes V_n
            \buildrel\mu'\over\rightarrow W.
\eeq Hence the map from $\a=\oph{\a}$ to $\tp{\a}$
arises naturally in the histories approach to
standard quantum theory. The important result from
this construction is that unlike the standard
representation with $\op{C}_\a$, no
      information about the homogeneous history $\oph{\a}$ is lost by
      representing it with the tensor product $\tp{\a}$ and unlike $\op{C}_\a$, the operator $\tp{\a}$ is a projection operator.

Hence, with the aid of the map $\th$, the operator
representation $\prod_{t\in\{t_1,\ldots
t_n\}}P(\H)_t$ of the space of homogeneous
histories $\U_{\{t_1,\ldots,t_n\}}$ with temporal
support $\{t_1,t_2,\ldots,t_n\}$ is embedded in
the space
$P(\otimes_{t\in\{t_1,\ldots,t_n\}}\H_t)$ of
projection operators on the Hilbert space
$\otimes_{t\in\{t_1,\ldots,t_n\}}\H_t$. The space
$P(\otimes_{t\in\{t_1,\ldots,t_n\}}\H_t)$ carries
the usual lattice structure of projection
operators and is therefore a natural model for the
space of history propositions based on homogeneous
histories with support $\{t_1,t_2,\ldots, t_n\}$.
In this model, history filters/homogeneous
histories are represented by homogeneous
projectors, and a general history proposition is
represented by an inhomogeneous projector. This
explains why the collection $\oph{\a}$ was refered
to earlier as a `homogeneous' history.

The decoherence functionals will be computed with
the aid of the map
$D:\otimes_{t\in\{t_1,\ldots,t_n\}}B(\H)\map
B(\H)$ defined by \beq
    D(\op{A}_1\otimes\op{A}_2\otimes\ldots\op{A}_n):=
    \op{A}_n(t_n)\op{A}_{n-1}(t_{n-1})\ldots\op{A}_1(t_1) \label{Def:D}
\eeq on homogeneous operators and then extended by
linearity. Thus, on a homogeneous history
$\a\in\U_{\{t_1,\ldots,t_n\}}$, the $C$-map is
defined by \beq
    \op{C}_\a:=D(\th(\a))                            \label{Def:C_a=D}
\eeq and then extended by linearity to the
appropriate set of inhomogeneous histories.

To incorporate arbitrary supports one needs to
collect together the operator algebras
$\otimes_{t\in s}B(\H)_t$ for all supports
$s\in\S$. The natural way of doing this is to use
an {\em infinite\/} tensor product of copies of
$B(\H)$.

   Let $\Omega$ denote a family of unit vectors in the Cartesian
product $\prod_{t\in\T}\H_t$ of copies of $\H$
labelled by the time values $t\in\T$; \ie
$t\mapsto\Omega_t$ is a map from $\T$ to the unit
sphere in $\H$. Then an infinite tensor product
$\otimes^\Omega_{t\in\T}B(\H)_t$ of operator
algebras $B(\H)$ is naturally associated with this
case. It is defined to be the weak closure (\ie
the closure in the weak operator topology) of the
set of all finite sums of functions from $\T$ to
$B(\H)$ that are equal to the unit operator for
all but a finite set of $t$-values. This
definition accomodates arbitrary temporal
supports, and the set of all projection operators
in $\otimes^\Omega_{t\in\T}B(\H)_t$ can be taken
as a model for the complete space $\UP$ of history
propositions in a standard Hilbert-space based,
quantum theory.

\subsection{The General Axioms for History Propositions}
 The axioms for the HPO approach to the consistent histories theory can be viewed as a more detailed
version of the original Gell-Mann and Hartle
axioms The general axioms and definitions are as
follows.

\noindent H1. {\em The space of history filters.}\
\ \noindent The fundamental ingredient in a theory
of histories is a space $\U$ of {\em history
filters\/}, or {\em possible universes\/}. This
space has the following structure.
\be
\item $\U$ is a partially-ordered set with a {\em unit\/}
      history filter $1$ and a {\em null\/} history filter $0$
      such that $0\leq\a\leq1$ for all $\a\in\U$.

\item $\U$ has a meet operation $\wedge$ which combines with the
      partial order $\leq$ to form a meet semi-lattice with unit $1$ so
      that $1\wedge\a=\a$ for all $\a\in\U$. The null history is
      absorptive in the sense that $0\wedge\a=0$ for all $\a\in\U$.

\item $\U$ is a partial semi-group with composition law
      denoted $\circ$. If $\a,\b\in\U$ can be combined to give
      $\a\circ\b\in\U$ we say that $\b$ {\em follows\/} $\a$, or
      $\a$ {\em preceeds\/} $\b$, and write $\a\lhd\b$. The $\circ$ and
      $\wedge$ laws are compatible in the sense that if $\a\circ\b$ is
      defined then it is equal to $\a\wedge\b$.

\item The null and unit histories can always be combined with any
      history filter $\a$ to give
      \beqa
        \a\circ 1&=&1\circ\a=\a         \\
        \a\circ 0&=&0\circ\a=0.
      \eeqa
\ee

\noindent H2. {\em The space of temporal
supports.}\ \ \noindent Any quasi-temporal
properties of the system are encoded in a partial
semi-group $\S$ of supports with unit $*$. The
support space has the following properties.
\be
\item There is a homomorphism $\s:\U\map\S$ of partial semi-groups
      that assigns a support to each history filter. The support of
      $0$ and $1$ is defined to be $*\in\S$.

\item {A history filter $\a$ is {\em nuclear\/} if it has no non-trivial
      decomposition of the form $\a=\b\circ\g$ with $\b,\g\in\U$; a
      temporal support $s$ is {\em nuclear\/} if it has no non-trivial
      decomposition of the form $s=s_1\circ s_2$ with $s_1,s_2\in\S$.
      Nuclear supports are the analogues of points of time; nuclear
      history filters are the analogues of single-time propositions.

      A decomposition of $\a\in\U$ as
      $\a=\a^1\circ\a^2\circ\ldots\circ\a^N$ is {\em irreducible\/} if
      the constituent history filters $\a^i\in\U$, $i=1,2,\ldots,N$ are
      all nuclear.

      A {\em resolution\/} of the semi-group homomorphism $\s:\U\map\S$
      is a chain of semigroups $\U_{i}$ and semi-group homomorphisms
      $\s_i$ so that $\s$ factors as the composition
      \beq
            \U{\buildrel\s_0\over\rightarrow}\U_1
              {\buildrel\s_1\over\rightarrow}\U_2
              {\buildrel\s_2\over\rightarrow}\ldots
              {\buildrel\s_{k-1}\over\rightarrow}\U_k
              {\buildrel\s_k\over\rightarrow}\S.
      \eeq
      }
\ee

\noindent H3. {\em The space of history
propositions.}\ \ \noindent The space $\U$ of
history filters is embedded in a larger space
$\UP$ of {\em history propositions\/}. This space
is an ortho-complemented lattice with a structure
that is consistent with the semi-lattice structure
on the subspace $\U$. One may also require the
lattice to be countably complete, or even
complete, depending on its cardinality. In
addition:
\be
\item The space $\UP$ can be generated from $\U$ by the
      application of a finite (or, perhaps, countably infinite) number
      of $\neg$, $\vee$ and $\wedge$ lattice operations. This captures
      the idea that elements of $\UP$ represent propositions `about'
      history filters (\ie about possible universes).

\item {An important role may be played by representations of the
      partial semi-group $\S$ in the automorphism group of the lattice
      $\UP$.

      Any representation of the ortho-complemented lattice $\UP$ by
      projection operators on a Hilbert space is called an {\em
      HPO quantisation\/} of the system.

       Two history propositions $\a$ and $\b$ are said to be {\em
      disjoint\/}, denoted $\a\perp\b$, if $\a\leq\neg\b$. A set of
      history propositions $\{\a^1,\a^2,\ldots,\a^N\}$ is {\em exclusive\/}
      if its elements are pairwise disjoint. It is {\em exhaustive\/} (or
      {\em complete\/}) if $\a^1\vee\a^2\vee\ldots\vee\a^N=1$.
      Countable sets (\ie with $N=\infty$) are permitted where
      appropriate.
      }
\ee It should be noted that the definitions above
are the direct HPO analogues of the corresponding
ideas introduced earlier in the context of the
Gell-Mann and Hartle axioms for consistent
histories.

\noindent H4. {\em The space of decoherence
functionals.}\ \ \noindent A {\em decoherence
functional\/} is a complex-valued map of pairs of
history propositions; the set of all such maps is
denoted $\cal D$. There may be a natural topology
on $\UP$ such that each decoherence functional
$d\in{\cal D}$ is a {\em continuous\/}
\footnote{Such a condition holds in standard
quantum theory because $\op{A}\mapsto
\tr(\op{A}\op{B})$ is a weakly continuous function
on bounded subsets of $B(\H)$ for each trace-class
operator $\op{B}$.} function of its arguments. Any
decoherence functional has the following
properties:
\be
\item The `inner-product' type conditions:
      \bi
      \item {\em Hermiticity\/}:\ \ $d(\a,\b)=d(\b,\a)^*$ for all
            $\a,\b\in\UP$.
      \item {\em Positivity\/}:\ \ $d(\a,\a)\ge0$ for all $\a\in\UP$.
      \item {\em Null triviality\/}:\ \ $d(0,\a)=0$ for all $\a\in\UP$.
      \ei

\item Conditions related to the potential probabilistic interpretation:
      \bi
      \item {\em Additivity\/}:\ \ if $\a\perp\b$ are general history
            propositions then, for all $\g\in\UP$,
            $d(\a\vee\b,\g)=d(\a,\g)+d(\b,\g)$.
      \item {\em Normalisation\/}: $d(1,1)=1$.
       \ei
\ee

\noindent H5. The (tentative) physical
interpretation of these axioms is the same as that
of the Gell-Mann and Hartle axioms, \ie the
diagonal element $d(\a,\a)$ is interpreted as the
probability of the history proposition $\a$ being
`true' when $\a$ is part of a consistent set. If
this is not the case, no direct physical meaning
is ascribed to the real number $d(\a,\a)$.

To summarise, in the HPO theory every history
proposition is represented as a projection
operator on a certain Hilbert space. This provides
valuable clues about the possible lattice
structure on $\UP$ in the general case and
suggests the existence of novel concepts. The
collection $\UP$ of all history propositions in a
general history theory can be equipped with a
lattice structure that is similar in some respects
to the lattice of propositions in standard quantum
logic. Any quasi-temporal properties of the theory
are coded in the space $\S$ of supports associated
with the subspace $\U$ of history filters. A
Boolean lattice would correspond to a history
version of a classical theory, and
quantum-mechanical superselection rules would
arise in the usual way via the existence of a
non-trivial center for the lattice $\UP$.

\chapter{Continuous Time in the History Projection Operator Theory}

\section{Introduction}
The introduction of a {\em continuous\/} time
clearly poses difficulties for any approach to the
consistent-history theory: in the class-operator
scheme one has to define continuous products of
projection operators. In the HPO approach, the
difficulty is to define a continuous {\em
tensor\/} product of projection operators.

In the original construction of the
continuous-time histories by Isham and Linden
\cite{IL95} the problem was resolved by exploiting
the existence of continuous tensor products of
coherent states. However, several interesting
issues were sidestepped in the process. For
example, the projectors onto coherent states do
not have a clear physical interpretation.
\par
In what follows, we will re-address the question
of continuous time in the HPO theory, inclining
towards a more physically-motivated construction.
As in \cite{IL95}, the starting point is the {\em
history group\/}: a history-analogue of the
canonical group used in standard quantum
mechanics. The key idea is that a unitary
representation of the history group leads to a
self-adjoint representation of its Lie algebra,
the spectral projectors of which are to be
interpreted as propositions about the histories of
the system. Thus we employ a history group whose
associated projection operators represent
propositions about continuous-time histories. As
we shall see, it will transpire that the history
algebra for one-dimensional quantum mechanics is
{\em infinite\/} dimensional---in fact, it is
isomorphic to the canonical commutation algebra of
a standard quantum {\em field\/} theory in one
spatial dimension. This suggests that it might be
profitable to study the history theory using tools
that are normally employed in quantum field
theory. In \cite{1}, we showed that the physically
appropriate representation of the history algebra
can be selected by requiring the existence of
operators that represent propositions about the
time-averaged values of the energy. The Fock space
thus constructed can be related to the notion of a
continuous tensor product as used in \cite{IL95},
thus establishing the link with the idea of
continuous temporal logic.

\section{The Choice of Time as a Continuous Parameter}

Most discussions of the consistent-histories
formalism have involved histories defined at a
finite set of discrete time points. However, it is
important to extend this to include a continuous
time variable, especially for potential
applications to quantum field theory and quantum
gravity.

As was mentioned in the Introduction, temporal
logic is a structure that ought to be of
particular importance in any approach to
consistent-histories theory. Indeed, this is one
of the key features of the HPO theory. In normal,
single-time quantum mechanics, a statement about a
physical quantity is represented by a projector on
the standard Hilbert space. Likewise in HPO, a
history (a {\em temporal\/} statement about
properties of the system), is represented by a
genuine projection operator (a tensor product of
projectors), on the tensor product of copies of
the standard Hilbert space. The one effect of
introducing the temporal logic by using tensor
products, is the fact that, there is no natural
way of generating a time translation from one time
slot to another. As we shall see, this situation
changes when we consider the {\em the continuous
limit\/} of such tensor products, in which time
now appears uniformly in a continuous way.

However, even after having made this step, one is
still involved in the use of quantities
(specifically, certain projectors in the history
quantum space)that do not have a clear physical
meaning. For this, Isham and Linden \cite{IL95}
introduced the history group for discrete-time
histories, and hence made a significant connection
in which the spectral projectors of the history
Lie algebra represent propositions about phase
space observables of the system. The next crucial
step is to introduce a continuous time variable by
introducing a delta function in the description of
its history commutation relations (at unequal
times).

As an immediate consequence, an intruiging feature
of the HPO theory appears; that all interesting
history propositions are about {\em
time-averaged\/} physical quantities. In other
words, the physical quantities in HPO are time
averaged and they cannot be defined at {\em sharp
moments in time\/}. Whether or not one considers
this to be a more natural way to identify physical
observables is, to a certain extent, a matter of
opinion. However, it is worth emphasising that, in
quantum field theories, only after proper
spacetime averaging (the analogue situation of the
history quantum mechanics time averaging) do the
operator-valued distributions correspond to
physical observables.

Hence, one can argue that a theory in which the
fundamental elements are time-averaged quantities
by construction---without at the same time
contradicting the standard quantum theory
treatment---generates an interest that should be
exploited.

Finally, it is worth adding that, to some extent,
it is a matter of personal opinion, whether or not
time should be regarded as a continuum. But then,
following a similar reasoning, it is natural to
expect the physical quantities not to be defined
at sharp moments of time. Hence the time-averaged,
or (field theoretic) spacetime-averaged
observables emerge from the physical
interpretation of the theory, rather than purely
from mathematical necessity.

\section{The History Space}

\subsection{The History Group}
We start by considering the HPO version of the
quantum theory of a particle moving on the real
line $\mathR$. As explained before, the history
proposition ``$\alpha_{t_1}$ is true at time
$t_1$, and then $\alpha_{t_2}$ is true at time
$t_2$, and then \ldots, and then $\alpha_{t_n}$ is
true at time $t_n$'' is represented by the
projection operator
$\alpha_{t_1}\otimes\alpha_{t_2}\otimes\cdots\otimes\alpha_{t_n}$,
on the $n$-fold tensor product ${\cal V}_n = {\cal
H}_{t_1}\otimes{\cal
H}_{t_2}\otimes\cdots\otimes{\cal H}_{t_n}$, of
$n$-copies of the Hilbert-space $\cal H$, of the
canonical theory. Since $\cal H$ carries a
representation of the Heisenberg-Weyl group with
Lie algebra
\begin{equation}
    {[\,}x,\, p\,] = i\hbar,                                \label{CCR}
\end{equation}
the Hilbert space ${\cal V}_n$ is expected to
carry a unitary representation of the $n$-fold
product group whose generators satisfy
\begin{eqnarray}
    {[\,}x_k,\,x_m\,]&=& 0                  \label{discreteHWxx} \\
    {[\,}p_k,\,p_m\,]&=& 0                  \label{discreteHWpp} \\
    {[\,}x_k,\,p_m\,]&=& i\hbar\delta_{km}  \label{discreteHWxp}
\end{eqnarray}
with $k,m = 1,2,\ldots,n$. Thus the Hilbert space
${\cal V}_n$ carries a representation of the
`history group' whose Lie algebra is defined to be
that of Eqs.\
(\ref{discreteHWxx})--(\ref{discreteHWxp}).
However, we can also turn the argument around and
{\em define\/} the history version of $n$-time
quantum mechanics by starting with Eqs.\
(\ref{discreteHWxx})--(\ref{discreteHWxp}). In
this approach, ${\cal V}_n$ arises as a
representation space for Eqs.\
(\ref{discreteHWxx})--(\ref{discreteHWxp}), and
tensor products
$\alpha_{t_1}\otimes\alpha_{t_2}\otimes\cdots\otimes\alpha_{t_n}$
that correspond to sequential histories about the
values of position or momentum (or linear
combinations of them) are then elements of the
spectral representations of this Lie algebra.

    We shall employ this approach to discuss continuous-time
histories. Thus, motivated by Eqs.\
(\ref{discreteHWxx})--(\ref{discreteHWxp}), we
start with the history-group whose Lie algebra
(referred to in what follows as the `history
algebra', or HA for short) is \cite{IL95}
\begin{eqnarray}
{[\,}x_{t_1},\,x_{t_2}\,] &=&0 \label{ctsHWxx} \\
{[\,}p_{t_1},\,p_{t_2}\,] &=&0 \label{ctsHWpp} \\
{[\,}x_{t_1},\,p_{t_2}\,]
&=&i\hbar\tau\delta(t_1-t_2)\label{ctsHWxp}
\end{eqnarray}
where $-\infty\leq t_1,\,t_2\leq\infty$ ; the
constant $\tau$ has dimensions of {\em time\/}
\cite{16}. Note that these operators are in the
{\em Schr\"odinger\/} picture: they must not be
confused with the Heisenberg-picture operators
$x(t), p(t)$ of normal quantum theory.

The choice of the Dirac delta function in the
right hand side of Eq.\ (\ref{ctsHWxp}), instead
of the Kronecker delta function that seems more
natural in dealing with Schr\"odinger picture
operators, is closely associated with the
requirement for treating time as a continuous
variable. As emphasised earlier, one consequence
is the fact that the observables cannot be defined
at sharp moments of time but rather as
time-averaged quantities.

    An important observation is that Eqs.\
(\ref{ctsHWxx})--(\ref{ctsHWxp}) are
mathematically the same as the canonical
commutation relations of a quantum {\em field \/}
theory in one space dimension:
\begin{eqnarray}
{[\,}\phi(x_1),\phi(x_2)\,]&=&0 \label{1Dphiphi}\\
{[\,}\pi(x_1),\pi(x_2)\,]&=&0 \label{1Dpipi}\\
{[\,}\phi(x_1),\pi(x_2)\,]&=&i\hbar\delta(x_1-x_2).
                                            \label{1Dphipi}
\end{eqnarray}

This analogy will be exploited fully in this
chapter. For example, the following two issues
arise immediately. Firstly---to be mathematically
well-defined---equations of the type Eqs.\
(\ref{ctsHWxx})--(\ref{ctsHWxp}) must be smeared
with test functions to give
\begin{eqnarray}
    {[\,}x_f,\,x_g\,]&=&0                   \label{SmearedCtsHWxx}  \\
    {[\,}p_f,\,p_g\,]&=&0                   \label{SmearedCtsHWpp}  \\
    {[\,}x_f,\,p_g\,]&=&i\hbar \tau\int_{-\infty}^\infty f(t)g(t)\,dt,
                                            \label{SmearedCtsHWxp}
\end{eqnarray}
which leads at once to the question of which class
$s$ of test functions to use. The minimal
requirement for the right hand side of Eq.\
(\ref{SmearedCtsHWxp}) to make sense is that $s$
must be a linear subspace of the space
$L^2(\mathR, dt)$ of square integrable functions
on $\mathR$. For the moment we shall leave $s$
unspecified beyond this.

    The second issue is concerned with finding the physically
appropriate representation of the HA Eqs.\
(\ref{SmearedCtsHWxx})--(\ref{SmearedCtsHWxp}),
bearing in mind that infinitely many unitarily
inequivalent representations are known to exist in
the analogous case of Eqs.\
(\ref{1Dphiphi})--(\ref{1Dphipi}). Note that this
problem does not arise in standard quantum
mechanics, or in the history version of quantum
mechanics with propositions defined at a finite
number of times, since---by the Stone-von Neumann
theorem---there is a unique representation of the
corresponding algebra up to unitarily equivalence.

    Of course, the physically appropriate Fock representation in the
 histories formalism is expected to involve some
type of continuous tensor product; this was the
path followed in \cite{IL95}. On the other hand,
according to a famous paper by Araki
\cite{Araki60},in standard quantum field theory,
the requirment that the Hamiltonian exists as a
proper self-adjoint operator is sufficient to
select a unique representation; for example, the
representations appropriate for a free boson field
with different masses are unitarily inequivalent.
In our case, this suggests that the appropriate
representation of the algebra Eqs.\
(\ref{SmearedCtsHWxx})--(\ref{SmearedCtsHWxp})
should be chosen by requiring the existence of
operators that represent history propositions
about (time-averaged) values of the energy. As we
shall see, this is indeed the case.

Before we exploit this statement further, it is
useful to briefly review the original work for
finding the representation space of the HA
\cite{IL95}.

\subsection{ The History Group Representation on $\Vcts$ }

From the perspective of the history theory the
physically appropriate representation space is
expected to involve some type of continuous tensor
product. Indeed in \cite{IL95}, Isham and Linden
showed how the requirement of representing the
history group on a Hilbert space (which we will
denote $\Vcts$) leads to a continuous tensor
product of copies of the standard Hilbert space
$L^2(\mathR, dx)$.

The starting point is the fact that the
representation space of the single-time Weyl group
Eq.\ (\ref{CCR}) can be written as a Fock space $
\exp{\cal{T}}$ where $\cal{T}\simeq \mathC$; hence
in order to find the representation of the history
group Eqs.\
(\ref{SmearedCtsHWxx})--(\ref{SmearedCtsHWxp}),
one proceeds by using Fock construction
techniques. This involves taking a
complexification of the space of real test
functions $L^2(\mathR, dx)$ used to smear the
generators of the history algebra Eqs.\
(\ref{SmearedCtsHWxx})--(\ref{SmearedCtsHWxp})
\beq
   {L_{\mathC}^2(\mathR, dx)} \cong {L_{\mathR}^2(\mathR, dx)} \oplus {L_{\mathR}^2(\mathR, dx)}
\eeq
where we have chosen the complexified space
to be $\cal{T}= \mathC$.

Each Fock space carries a special class of states,
the normalised coherent states defined as \beq
    \ket z \eqdef e^{-\half |z|^2 +za^\dagger}\ket 0,
\eeq and satisfy $\braket zw = e^{-\half
|z|^2-\half |w|^2 +z^*w}$. The idea of coherent
states is a necessary mathematical tool in the
construction of continuous tensor products. They
are related to the elements $\ket {\exp z} $ of
the Fock space $\exp{\mathC}$ by $\ket{\exp z}=
e^{\half|z|^2}\ket{z}$, and the Hilbert space
$\exp\,\mathC$ is isomorphic to
$L_{\mathC}^2(\mathR)$ via \beqa
    \exp\,\mathC &\simeq& L_{\mathC}^2 (\mathR,dx) \label{ExpC}                \\
    \ket{\exp z} &\mapsto& \braket{x}{\exp z}=(2\pi)^{-{1\over 4}}
            e^{zx-\half z^2 -{1\over 4}x^2}.                                \nn
\eeqa

On the other hand, we consider the continuous
tensor product of a one-parameter family of $ t
\longmapsto {\cal{H}}_t $ of standard Hilbert
spaces. In general, one may try to define the
inner product as \beq {\braket{{\ot_t u_t}}{{\ot_t
v_t}}}_{\ot_t\H_t} \eqdef
e^{\int_{-\infty}^\infty\log {\inner{{u_t}}{
v_t}}_{\H_t}\,dt} \eeq This is intended to be the
continuous analogue of the inner product between
discrete tensor products of vectors \beq
    \braket{u_1\ot u_2\ot\cdots\ot u_n}{v_1\ot v_2\ot\cdots\ot v_n}
    \eqdef \prod_{i=1}^n\inner{u_i}{v_i}\equiv
        e^{\sum_{i=1}^n\log\inner{u_i}{v_i}}.
\eeq If $\H_t$ is an exponential Hilbert space
$\H_t=\exp \K_t$, then the construction works
since \beq {\braket{{\exp \phi_t}}{{\exp
\psi_t}}}_{e^{\K_t}} {=}
e^{\inner{\phi_t}{\psi_t}_{\K_t}} \eeq and so the
definition of the scalar product on the continuous
tensor product of copies of $\exp\K_t$ as \beq
{\braket{{\ot_t\exp
\phi_t}}{{\ot_t\exp\psi_t}}}_{\ot_t
e^{\K_t}}\eqdef e^{\int_{-\infty}^\infty
{\inner{{\phi_t}}{{\psi_t}}}_{\K_t}\,dt} \eeq is
well-defined.

Furthermore, the scalar product
$\int_{-\infty}^\infty
{\inner{{\phi_t}}{{\psi_t}}}_{\K_t}\,dt$ is the
inner product on the direct integral Hilbert space
$\int^\oplus \K_t$, hence we can write \beq
{\braket{{\ot_t\exp \phi_t}}{{\ot_t\exp
\psi_t}}}_{\ot_t e^{\K_t}} {=}
 {\ebraket{\phi}{\cdot}{\psi}{\cdot}}_{\exp\int^\oplus \K_t}
\eeq In fact, there exists the useful isomorphism
\beqa
    \ot_t \exp \K_t &\simeq& \exp\int^\oplus\K_t \label{ExpIso} \\
    \ot_t\ket{\exp\phi_t}&\mapsto& \ket{\exp\phi(\cdot)}                    \nn
\eeqa The existence of yet another isomorphism
between the Hilbert space $L^2(\mathR)$ and the
direct integral $\int^\oplus \mathC_t\,dt$ via
\beqa
    \int^\oplus \mathC_t\,dt &\simeq& L^2(\mathR,dt) \\
    \int^\oplus w_t\,dt &\mapsto& w(\cdot)
\eeqa links together the previous results with the
Fock space $ \exp\,\mathC_t\simeq L^2_t(\mathR)$
that represents the history group, {\em i.e.\/}
the continuous-time Weyl group.

Indeed, for the special case $ \K_t = {\mathC}_t
$, we summarise the previous isomorphisms as \beqa
    \exp\,\mathC_t\simeq L^2_t(\mathR)  \\
    \otimes_t\exp\,\mathC_t\simeq\exp\int^\oplus \mathC_t\,dt  \\
    \exp(L^2(\mathR))\simeq\exp\int^\oplus \mathC_t\,dt
\eeqa to conclude that \beq
    \Vcts\eqdef\ot_t\lbra L^2_t(\mathR)\rbra
        \simeq \exp\lbra L^2(\mathR,dt)\rbra.
\eeq

Hence, the Fock space $\exp(L^2(\mathR))$ on which
the history algebra is naturally represented, is
isomorphic to the space $\exp\int^\oplus
\mathC_t\,dt$, the continuous tensor product of
copies of the standard Hilbert space.

\subsection{The Hamiltonian Algebra}
Having found the representation space of the
history group from a mathematical perspective, we
now turn to explore a more physically meaningful
way of uniquely selecting the representation space
of the history algebra.

We return to the crucial observation---for the
construction of the theory---that there exists a
strong resemblance between the history algebra
Eqs.\ (\ref{discreteHWxx})--(\ref{discreteHWxp}),
and the canonical commutation relations of a
quantum field theory in one space dimension
\begin{eqnarray}
{[\,}\phi(x_1),\phi(x_2)\,]&=&0 \label{1Dphiphi}\\
{[\,}\pi(x_1),\pi(x_2)\,]&=&0 \label{1Dpipi}\\
{[\,}\phi(x_1),\pi(x_2)\,]&=&i\hbar\delta(x_1-x_2).
                                            \label{1Dphipi}
\end{eqnarray}

We start with the ubiquitous example of the
one-dimensional, simple harmonic oscillator with
Hamiltonian
\begin{equation}
    H={p^2\over 2m}+{m\omega^2\over 2}x^2.
\end{equation}
As we have seen, the na{\"\i}ve idea behind the
HPO theory is that to each time $t$ there is
associated a Hilbert space ${\cal H}_t$ that
carries propositions appropriate to that time (the
`na{\"\i}vety' refers to the fact that, in a {\em
continuous\/} tensor product
$\otimes_{t\in\mathR}{\cal H}_t$, the individual
Hilbert spaces ${\cal H}_t$ do not strictly exist
as subspaces; this is related to the need to smear
operators). Thus we expect to have a one-parameter
family of operators
\begin{equation}
H_t:={p_t^2\over 2m}+{m\omega^2\over 2}x_t^2
\label{Def:Ht}
\end{equation}
that represent the energy at time $t$.

    As it stands, the right hand side of Eq.\ (\ref{Def:Ht}) is not
well-defined, just as in normal canonical quantum
field theory it is not possible to define products
of field operators at the same spatial point.
However, the commutators of $H_t$ with the
generators of the HA can be computed formally as
\begin{eqnarray}
&&{[\,}H_t,\,x_s\,]=-{i\hbar\over m}\delta(t-s)p_s  \label{[Htxs]}\\
&&{[\,}H_t,\,p_s\,]=i\hbar m\omega^2\delta(t-s)x_s  \label{[Htps]}\\
&&{[\,}H_t,\,H_s\,]=0                               \label{[HtHs]}
\end{eqnarray}
and are the continuous-time, history analogues of
the familiar result in standard quantum theory:
\begin{eqnarray}
    {[\,}H,\,x\,]&=&-{i\hbar\over m }p              \label{[Hx]}\\
    {[\,}H,\,p\,]&=&i\hbar m\omega^2 x.             \label{[Hp]}
\end{eqnarray}

    In standard quantum theory, the spectrum of the Hamiltonian
operator can be computed directly from the algebra
of Eqs.\ (\ref{[Hx]})--(\ref{[Hp]}) augmented with
the requirement that the underlying representation
of the canonical commutation relations Eq.\
(\ref{CCR}) is irreducible. This suggests that we
try to {\em define\/} the history theory by
requiring the existence of a family of operators
$H_t$ that satisfy the relations Eqs.\
(\ref{[Htxs]})--(\ref{[HtHs]}) and where the
representation of the canonical history algebra
Eqs.\ (\ref{ctsHWxx})--(\ref{ctsHWxp}) is
irreducible. More precisely, we augment the HA
with the algebra (in semi-smeared form)
\begin{eqnarray}
    &&{[\,}H(\chi),\,x_t\,]=-{i\hbar\over m}\chi(t)p_t
                                                \label{[Hchixt]}\\
    &&{[\,}H(\chi),\,p_t\,]=i\hbar m\omega^2\chi(t)x_t
                                                \label{[Hchipt]}\\
    &&{[\,}H(\chi_1),\,H(\chi_2)\,]=0           \label{[Hchi1Hchi2]}
\end{eqnarray}
where $H(\chi)$ is the history energy-operator,
time averaged with the function $\chi$;
heuristically,
$H(\chi)=\int_{-\infty}^{\infty}dt\, \chi(t) H_t$.

    It is useful to integrate these equations in the following
sense. {\em If\/} self-adjoint operators $H(\chi)$
exist satisfying Eqs.\
(\ref{[Hchixt]})--(\ref{[Hchi1Hchi2]}), we can
form the unitary operators $e^{iH(\chi)/\hbar}$,
and these satisfy
\begin{eqnarray}
    &&e^{iH(\chi)/\hbar}\,x_t\,e^{-iH(\chi)/\hbar}
    =\cos[\omega\chi(t)]x_t +{1\over m\omega}\sin[\omega\chi(t)]p_t
                                \label{UxtU}\\
    &&e^{iH(\chi)/\hbar}\,p_t\,e^{-iH(\chi)/\hbar}
        =-m\omega\sin[\omega\chi(t)]x_t+\cos[\omega\chi(t)]p_t.
                                                        \label{UptU}
\end{eqnarray}
However, it is clear that the right hand side of
Eqs.\ ({\ref{UxtU})--({\ref{UptU}) defines an {\em
automorphism\/} of the canonical history algebra
Eqs.\ (\ref{ctsHWxx})--(\ref{ctsHWxp}). Thus the
task in hand can be rephrased as that of finding
an irreducible representation of the HA in which
these automorphisms are unitarily implementable:
the self-adjoint generators of the corresponding
unitary operators will then be the desired
time-averaged energy operators $H(\chi)$ [strictly
speaking, weak continuity is also necessary, but
this poses no additional problems in the cases of
interest here].

\subsection{The Fock Representation}
It is natural to contemplate the use of a Fock
representation of the HA since this plays such a
central role in the analogue of a free quantum
field in one spatial dimension. To this end, we
start by defining the `annihilation operator'
\begin{equation}
    b_t:=\sqrt{{m\omega\over2\hbar}}x_t+
            i\sqrt{{1\over 2m\omega\hbar}}p_t   \label{Def:bt}
\end{equation}
in terms of which the HA
(\ref{ctsHWxx})--(\ref{ctsHWxp}) becomes
\begin{eqnarray}
    &&{[\,}b_t,\,b_s\,]=0                           \label{[btbs]}\\
    &&{[\,}b_t,\,b_s^\dagger\,]=\delta(t-s).        \label{[btbsdag]}
\end{eqnarray}
Note that
\begin{equation}
    \hbar\omega b_t^\dagger b_s={1\over 2m}p_tp_s
        +{m\omega^2\over 2}x_tx_s-{\hbar\omega\over2}\delta(t-s)
\end{equation}
which suggests that there exists an additively
renormalised version of the operator $H_t$ in Eq.\
(\ref{Def:Ht}) of the form $\hbar\omega
b_t^\dagger b_t$. In turn, this suggests strongly
that a Fock space based on Eq.\ (\ref{Def:bt})
should provide the operators we seek.

    To make this explicit we recall that the bosonic Fock space
${\cal F}[{\cal H}]$ associated with a Hilbert
space $\cal H$ is defined as
\begin{equation}
    {\cal F}[{\cal H}]:=\mathC\oplus {\cal H}\oplus
            ({\cal H}\otimes_S{\cal H})\oplus\cdots
\end{equation}
where ${\cal H}\otimes_S{\cal H}$ denotes the
symmetrised tensor product of $\cal H$ with
itself. Any unitary operator $U$ on the
`one-particle' space $\cal H$ gives a unitary
operator $\Gamma(U)$ on ${\cal F}[{\cal H}]$
defined by
\begin{equation}
    \Gamma(U):=
        1\oplus U\oplus (U\otimes U)\oplus\cdots    \label{Def:G(U)}
\end{equation}
Furthermore, if $U=e^{iA}$ for some self-adjoint
operator $A$ on $\cal H$, then
$\Gamma(U)=e^{id\Gamma(A)}$ where
\begin{equation}
    d\Gamma(A):=0\oplus A\oplus (A\otimes1+1\otimes A)\oplus\cdots.
                                                \label{Def:dGA}
\end{equation}

    The implications for us of these well-known constructions are as
follows. Consider the Fock space ${\cal
F}[L^2(\mathR,dt)]$ that is associated with the
Hilbert space $L^2(\mathR,dt)$ via the
annihilation operator $b_t$ defined in Eq.\
(\ref{Def:bt}); {\em i.e.,} the space built by
acting with (suitably smeared) operators
$b_t^\dagger$ on the `vacuum state' $\ket{0}$ that
satisfies $b_t\ket{0}=0$ for all $t\in\mathR$. The
equations Eq.\ (\ref{UxtU})--(\ref{UptU}) show
that, {\em if\/} it exists, the operator
$e^{iH(\chi)/\hbar}$ acts on the putative
annihilation operator $b_t$ as
\begin{equation}
    e^{iH(\chi)/\hbar}\, b_t\, e^{-iH(\chi)/\hbar} =
        e^{-i\omega\chi(t)}b_t.  \label{UbtU}
\end{equation}
However, thought of as an action on
$L^2(\mathR,dt)$, the operator $U(\chi)$ defined
by
\begin{equation}
        (U(\chi)\psi)(t):=e^{-i\omega\chi(t)}\psi(t)
\end{equation}
is unitary for any measurable function $\chi$.
Hence, using the result mentioned above, it
follows that in this particular Fock
representation of the HA the automorphism on the
right hand side of Eq.\ (\ref{UbtU}) {\em is\/}
unitarily implementable, and hence the desired
self-adjoint operators exist. Note that
$H(\chi)=\hbar\omega\, d\Gamma(\hat\chi)$, where
the self-adjoint operator $\hat\chi$ is defined on
$L^2(\mathR,dt)$ as
\begin{equation}
    (\hat\chi\psi)(t):=\chi(t)\psi(t).
\end{equation}

    In summary, we have shown that the Fock representation of the
HA Eqs.\ (\ref{ctsHWxx})--(\ref{ctsHWxp})
associated with the annihilation operator $b_t$ of
Eq.\ (\ref{Def:bt}) is such that there exists a
family of self-adjoint operators $H(\chi)$ for
which the algebra Eqs.\
(\ref{[Hchixt]})--(\ref{[Hchi1Hchi2]}) is
satisfied. This Fock space is the desired carrier
of the history propositions in our theory. Note
that, in this case, the natural choice for the
test function space $\phi\subseteq L^2(\mathR,dt)$
used in Eqs.\
(\ref{SmearedCtsHWxx})--(\ref{SmearedCtsHWxp}) is
simply $L^2(\mathR,dt)$ itself.

    The position history-variable $x_t$ can be written in terms of
$b_t$ and $b_t^\dagger$ as
\begin{equation}
    x_t=\sqrt{{\hbar\over2m\omega}}\left(b_t+b_t^\dagger\right)
\end{equation}
and has the correlation function
\begin{equation}
        \bra{0}x_t\,x_s\ket{0}={\hbar\over2m\omega}\delta(t-s).
                                \label{<xtxs>}
\end{equation}
Thus the carrier space of our history theory is
Gaussian white noise.

\section{The n-particle History Propositions}
The Fock-space construction produces a natural
collection of history propositions: namely, those
represented by the projection operators onto what,
in a normal quantum field theory, would be called
the `$n$-particle states'. To see what these
correspond to physically in our case we note first
that a $\delta$-function normalised basis for
${\cal F}[L^2(\mathR,dt)]$ is given by the vectors
$\ket{0}$, $\ket{t_1}$, $\ket{t_1,t_2}$, \ldots
where $\ket{t_1}:=b_{t_1}^\dagger\ket{0}$,
$\ket{t_1,t_2}:=b_{t_1}^\dagger
b_{t_2}^\dagger\ket{0}$, {\em etc\/} (of course,
properly normalised vectors are of the form
$\ket{\phi}:=b_\phi^\dagger\ket{0}$ {\em etc} for
suitable smearing function $\phi$). The physical
meaning of the projection operators of the form
$\ketbra{t}{t}$ (or, more rigorously,
$\ketbra{\phi}{\phi}$),
$\ketbra{t_1,t_2}{t_1,t_2}$, {\em etc\/}, can be
seen by studying the equations
\begin{eqnarray}
    &&H(\chi)\ket{0}=0 \\ &&H(\chi)\ket{t}=\hbar\omega\chi(t)\ket{t}
\\ &&H(\chi)\ket{t_1,t_2}=\hbar\omega[\chi(t_1)+\chi(t_2)]
\ket{t_1,t_2}
\end{eqnarray}
or, in totally unsmeared form,
\begin{eqnarray}
    &&H_t\ket{0}=0                                      \\
    &&H_t\ket{t_1}=\hbar\omega\delta(t-t_1)\ket{t_1}    \\
    &&H_t\ket{t_1,t_2}=\hbar\omega[\delta(t-t_1)+\delta(t-t_2)]
                        \ket{t_1,t_2}.
\end{eqnarray}

    It is clear from the above that, for example, the projector
$\ketbra{t_1,t_2}{t_1,t_2}$ represents the
proposition that there is a unit of energy
$\hbar\omega$ concentrated at the time point $t_1$
and another unit concentrated at the time point
$t_2$. Note that
$H(\chi)\ket{t,t}=2\hbar\omega\chi(t)\ket{t,t}$,
and hence $\ketbra{t,t}{t,t}$ represents the
proposition that there are two units of energy
concentrated at the {\em single\/} time point $t$
(thus exploiting the Bose-structure of the
canonical history algebra!). This interpretation
of projectors like $\ketbra{t_1,t_2}{t_1,t_2}$ is
substantiated by noting that the time-averaged
energy obtained by choosing the averaging function
$\chi$ to be $1$ acts on these vectors as
\begin{eqnarray}
    &&\int_{-\infty}^\infty ds\, H_s\ket{t}=\hbar\omega\ket{t}\\
    &&\int_{-\infty}^\infty ds\, H_s\ket{t_1,t_2}=
            2\hbar\omega\ket{t_1,t_2}
\end{eqnarray}
and so on. This is the way in which the HPO
account of the simple harmonic oscillator recovers
the integer-spaced energy spectrum of standard
quantum theory.

    Finally, we note in passing that
\begin{equation}
    {1\over\hbar\omega}\int_{-\infty}^\infty ds\,
    sH_s\,\ket{t_1,t_2,\ldots,t_n}=
            (t_1+t_2+\cdots+t_n)\ket{t_1,t_2,\ldots,t_n}
\end{equation}
so that ${1\over\hbar\omega}\int_{-\infty}^\infty
ds\, sH_s$ acts as a `total time' or
`center-of-time' operator.

\section{The Extension to Three Dimensions}
The extension of the formalism above to a particle
moving in three spatial dimensions appears at
first sight to be unproblematic. The analogue of
the history algebra Eqs.\
(\ref{ctsHWxx})--(\ref{ctsHWxp}) is
\begin{eqnarray}
    {[\,}x^i_{t_1},\,x^j_{t_2}\,] &=&0 \label{ctsHWxx3D} \\
{[\,}p^i_{t_1},\,p^j_{t_2}\,] &=&0
\label{ctsHWpp3D} \\ {[\,}x^i_{t_1},\,p^j_{t_2}\,]
&=& i\hbar\delta^{ij}\delta(t_1-t_2)
                                    \label{ctsHWxp3D}
\end{eqnarray} $i,j=1,2,3$; while the formal expression Eq.\
(\ref{Def:Ht}) for the energy at time $t$ becomes
\begin{equation}
    H_t:={\underline{p}_t\cdot\underline{p}_t\over 2m}+
{m\omega^2\over
2}\underline{x}_t\cdot\underline{x}_t.
\label{Def:Ht3D}
\end{equation}

    It is straightforward to generalise the discussion above to this
situation and, in particular, to find a Fock
representation of Eqs.\
(\ref{ctsHWxx3D})--(\ref{ctsHWxp3D}) in which the
rigorous analogues of Eq.\ (\ref{Def:Ht3D}) exist
as properly defined self-adjoint operators.
However, an interesting issue then arises that has
no analogue in one-dimensional quantum theory.
Namely, we expect to have angular-momentum
operators whose formal expression is
\begin{equation}
    L^i_t:=\epsilon^i{}_{jk}x^j_tp^k_t          \label{Def:Lit}
\end{equation}
and whose commutators can be computed
heuristically as
\begin{equation}
{[\,}L^i_t,\,L^j_s\,]=i\hbar\epsilon^{ij}{}_k\delta(t-s)L^k_t.
                                                \label{[LitLjs]}
\end{equation}
Such operators $L^i_t$ can be constructed
rigorously using, for example, the method employed
for the energy operators $H_t$: viz., compute the
automorphisms of the canonical history algebra
that are formally induced by the angular-momentum
operators and then see if these automorphism can
be unitarily implemented in the given Fock
representation. However, the interesting
observation is that, even if this can be done
(which is the case, see below), this does not
guarantee in advance that the commutators in Eq.\
(\ref{[LitLjs]}) will be reproduced: in
particular, it is necessary to check directly if a
$c$-number {\em central extension\/} is present
since we know from other branches of theoretical
physics that algebras of the type in Eq.\
(\ref{[LitLjs]}) are prone to such anomalies.

An obvious technique for evaluating such a
commutator would be to define the angular momentum
operators by point-splitting in the form
\begin{equation}
        L^i_{t,\epsilon} := i\hbar\epsilon^{i}{}_{jk}
            (b^j_{t})^\dagger b^k_{t+\epsilon}
\end{equation}
so that the commutator in Eq.\ (\ref{[LitLjs]}) is
the analogue of an equal-time commutator in
standard quantum field theory, and the
point-splitting is the analogue of spatial point
splitting. It is then straightforward to compute
the commutators of these point-split operators and
take the limit $\epsilon\rightarrow 0$. The result
is the anticipated algebra Eq.\ (\ref{[LitLjs]}).

    However, in standard quantum field theory it is known that the
limit of the commutator has to be considered at
unequal times ({\em i.e.}, using
Heisenberg-picture operators), and that there is a
subtle relation between the two limits of the
times becoming equal and the spatial point
splitting tending to zero\cite{Jackiw}. Therefore,
in order to calculate correctly the commutator in
our case it seems appropriate to consider the
analogue of an unequal time commutator, namely
\begin{equation}
    [L^i_{\chi,t,\epsilon},\,L^j_{0,s,\epsilon}]
\end{equation}
where
\begin{equation}
    L^i_{\chi,t,\epsilon}:= i\hbar\epsilon^{i}{}_{jk}(b^j_{\chi,t})^\dagger
        b^k_{\chi,t+\epsilon},
\end{equation}
and where
\begin{equation}
         b^k_{\chi,t} := e^{iH(\chi)} b^k_t e^{-iH(\chi)}
         = e^{-i\omega\chi(t)} b^k_t
\end{equation}
is a time-averaged Heisenberg picture operator of
the type defined earlier.

    It is not difficult to show that
\begin{eqnarray}
        [L^i_{\chi,t,\epsilon},\,L^j_{0,s,\epsilon}] &&=
    -\hbar^2 e^{i\omega(\chi(t)-\chi(t+\epsilon))}\nonumber\\
&&\times\big[\delta(t-s+\epsilon)\left((b^j_{t})^\dagger b^i_{t+2\epsilon} -
\delta^{ij}(b^m_{t})^\dagger b^m_{t+2\epsilon}
\right)\nonumber\\
&&\qquad-\delta(t-s-\epsilon)
\left( (b^i_{t-\epsilon})^\dagger b^j_{t+\epsilon}
- \delta^{ij}(b^m_{t-\epsilon})^\dagger
b^m_{t+\epsilon} \right) \big]
\end{eqnarray}
and then, by evaluating the matrix element of the
commutator in the vacuum state, one sees that
there is no central extension in this case.
Furthermore, by considering the matrix element of
the commutator in general coherent states, one can
check that the limits of $\epsilon \rightarrow 0$
and $\chi\rightarrow 0$ are straightforward, and
that as long as the test functions are smooth, the
angular momentum generators do indeed satisfy the
heuristic commutator Eq.\ (\ref{[LitLjs]}) in the
limit.

\chapter{The Action Operator}

\section{Introduction}
In the previous chapter we showed that, for the example of a simple
harmonic oscillator in one dimension, the requirement of the
existence of the Hamiltonian operator---which represents
propositions about the time-averaged values of the energy
of the system---together with the
explicit relation between the Hamiltonian and the creation
and annihilation operators, uniquely selectes a particular Fock space
as the representation space of the history algebra
[1.2-1.4] on the history space ${\cal{V}}_{cts}$.

The history algebra generators $x_t$ and $p_t$ can be seen
heuristically as operators, (actually they are
operator-valued distributions on ${\cal{V}}_{cts}$), that for
each time label $t$, are defined on the Hilbert space
${\cal{H}}_t$. The question then arises if, and how, these
Schr\"odinger-picture objects with different time labels
are related: in particular, is there a transformation law
`from one Hilbert space to another'? One anticipates that
the analogue of this question in the context of a histories
treatment of a relativistic quantum field theory would be
crucial to showing the Poincar\'e invariance of the system. 
The main goal of the present chapter is to enhance the theory so as to have a clearer view of the time transformation issue. This will ultimately allow us to address the problem of the Poincar\'e covariance of a history version of quantum field theory \cite{In preparation}.

In the Hamilton-Jacobi formulation of Classical Mechanics,
it is the {\em action functional\/} that plays the role of
the generator of a canonical transformation of the system
from one time to another. Indeed, the Hamilton-Jacobi
functional $S$, evaluated on the realised path of the
system---{\em i.e.}, for a solution of the classical
equations of motion, under some initial conditions---is the
generating function of a canonical transformation which
transforms the system variables, for example position $x$ and momentum $p$, from an initial time $t=0$ to another time $t$. It is
therefore natural to investigate whether a quantum analogue
of the action functional exists for the HPO theory and what role it plays in regard to the concept of time evolution in the theory.

At this point, it is worth commenting on the fact that there is an interesting relation between the definition of the action operator in HPO, and the well known work by Dirac, on the Langrangian theory for quantum mechanics \cite{Dirac}. Motivated by the fact that---contrary to the Hamiltonian---the Langrangian method can be expressed relativistically on account of the action function being a relativistic invariant, Dirac tried to take over the $\it{ideas}$ of the classical Langrangian theory, albeit not the equations of the Langrangian theory {\em per se\/} \cite{Dirac}. In doing so, he showed that the transformation function $ {\braket{q}{Q}} $ that connects the two position representations---in which  $q$ (the position at time t) and $Q$ (the position at another time $\prime t$) are multiplicative operators---is the quantum analogue of $e^{\frac{i}{\hbar} S}$ where $S$ is the classical action functional. He also obtained the contact transformations of the classical action functional in quantum mechanics; from this work the path integral approach to quantum theory was eventualy developed.
We will show that in HPO, the quantum analogue of the action functional acts in a similar way: it is the generator of  time transformations in the sense that it relates the position and the momentum observables of the system at one time, with the position and momentum observables at another time, (as mentioned above, it resembles the canonical transformations generated by the Hamilton-Jacobi action functional).

In what follows, we prove the existence of the action
operator $S_{\kappa}$, using the same type of quantum field
theory methods that were used earlier to prove the existence of the
Hamiltonian operator $H_{\kappa}$.

\section{The Definition of the Action Operator}

In the Hamiltonian formalism for a classical system, the
action functional is defined as
\begin{equation}
  S_{\rm{cl}}:=  \int^{+\infty}_{-\infty}(p\dot{q}-H)dt  \label{klas}
\end{equation} 
where $q$ is the position, $p$ is the momentum and $H$ the
Hamiltonian of the system. Following the same line of
thought as we the one we used in the definition of the Hamiltonian algebra,
we want to find a representation of the history algebra in
which their exists a one-parameter family of operators
$S_{t}$---or better, their smeared form $S_{\lambda
,\kappa}$. Heuristically, we have
\begin{eqnarray}
   S_{t}&:=&(p_{t}\dot{x}_{t}- H_{t})     \\
   S_{\lambda,\kappa}&:=&\int^{+\infty}_{-\infty}
            (\lambda(t)p_{t}\dot{x}_{t}-\kappa(t)H_{t})dt
\end{eqnarray}
where $ S_{\lambda,\kappa} $ is the smeared action operator
with smearing functions $\lambda(t), \kappa(t)$. In order
to discuss the existence of an operator
$S_{\lambda,\kappa}$ we note that, {\em if\/} this operator
exists, the Hamiltonian algebra eqs.\ (\ref{[Hchixt]}---\ref{[Hchi1Hchi2]}), would be augmented in the form
\begin{eqnarray}
[\,S_{\lambda,\kappa}, x_{f}\,]&=&
        i\hbar({x}_{\frac{d}{dt}(\lambda f)} +
            \frac{p_{\kappa f}}{m})  \label{S1}   \\
{[}\, S_{\lambda,\kappa}, p_{f}\,]&=& i\hbar({p}_{\frac{d}{dt}(\lambda
f)} + m\omega x_{\kappa f})  \label{S2}   \\
 {[}S_{\lambda,\kappa} ,
H_{{\kappa}^{\prime}}\,]&=& i\hbar H_{\frac{d}{dt}(\lambda
\kappa^{\prime})} - \frac{i\hbar}{m}
\int^{\infty}_{-\infty} (\kappa^{\prime}(t) \frac{d}{dt} ({\lambda}
(t){p}_t^2))dt   \label{S3}   \\
 {[}\,S_{\lambda,\kappa} , S_{\lambda,\kappa}^{\prime}\,]&=&
i\hbar H_{\frac{d}{dt}( \lambda^{\prime} \kappa )} - i\hbar
H_{\frac{d}{dt}( \lambda \kappa^{\prime} ) } -\nonumber\\
\hspace{2cm}&&
 i\hbar \int^{\infty}_{-\infty}( [(\kappa(t)
\frac{d}{dt} {\lambda^{\prime}}(t))-( \kappa^{\prime}(t)
\frac{d}{dt} {\lambda}(t))]\frac{\frac{d}{dt}p_t^2}{m}) dt  \label{S4}
\end{eqnarray}

Although we have defined the action operator in a general
smeared form, in what follows we will mainly employ only
the case $\lambda(t)=1$ and $\kappa(t)=1$ that accords with
the expression for the classical action functional. This
choice of smearing functions poses no technical problems, provided we keep to the requirement that the
smearing functions for the position and momentum operators
are square-integrable functions. In particular, the
products of the smearing functions $f$ and $g$ in
eqs (\ref{S1}---\ref{S4}) with the test functions $\lambda(t)=1$ and
$\kappa(t)=1$ are still square-integrable.

\paragraph*{The Existence of the Action Operator in HPO.}
We now examine whether the action operator actually exists
in the Fock representation of the history algebra employed
in our earlier discussion in Chapter 3. Henceforward we choose
$\lambda(t)=1$. Then the formal commutation relations are
\begin{eqnarray}
 S_{\kappa}&:=& \int^{+\infty}_{-\infty}( p_{t}\dot{x}_{t}- \kappa(t)H_{t})dt     \\
  {[}\, S_{\kappa} , x_{f}\,]&=& i\hbar({x}_{\dot{f}} + \frac{p_{\kappa f}}{m})     \\
  {[}\, S_{\kappa} , p_{f}\,]&=& i\hbar({p}_{\dot{f}} + m\omega
            x_{\kappa f}) \\
   {[}\, S_{\kappa} , H_{{\kappa}^{\prime}}\,] &=& i\hbar
        H_{{\dot{\kappa}}^{\prime}} \\
   {[}\, S_{\kappa} , S_{{\kappa}^{\prime}}\,]&=& i\hbar
H_{\dot{\kappa}} -i\hbar H_{{\dot{\kappa}}^{\prime}}
\end{eqnarray}

A key observation is that {\em if\/} the operators
$e^{\frac{i}{\hbar}S_{\kappa}}$ existed they would produce
the history algebra automorphism
\begin{equation}
 e^{\frac{i}{\hbar}s S_{\kappa}} b_{t} e^{-\frac{i}{\hbar}s S_{\kappa}}=
e^{i\omega \int^{t+s}_{t} \kappa(t+s^{\prime}) ds^{\prime}}
  e^{s\frac{d}{dt}}b_{t}  \label{Sauto}
\end{equation}
or, in the more rigorous smeared form
\begin{equation}
 e^{\frac{i}{\hbar}sS_{\kappa}} b_{f}
 e^{\frac{i}{\hbar}sS_{\kappa}}= b_{{\Sigma}_{s} f}
\end{equation}
where the unitary operator $\Sigma_s$ is defined on
${L}^{2}({\mathbf{R}})$ by
\begin{equation}
  ({\Sigma}_{s}\psi)(t):=
        e^{-i\omega \int^{t+s}_{t} \kappa(t+s^{\prime})
                ds^{\prime}} \psi(t+s).\label{Def:Sigma}
\end{equation}

As we explained in the previous chapter, an important property of the Fock construction
states that if
$e^{i s A }$ is a unitary operator on the Hilbert space
$L^{2}({\mathbf{R}})$, there exists a unitary
operator $\Gamma(e^{i s A })$ that acts on the
exponential Fock space
${\mathcal{F}}(L^{2}({\mathbf{R}}))$ in such a
way that
\begin{equation}
 \Gamma(e^{i s A }) b^{\dagger}_{f}
{\Gamma(e^{i s A })}^{-1} =
{b^{\dagger}}_{e^{i s A }f}.   \label{prop1}
\end{equation}
In addition, there exists a self-adjoint operator $d\Gamma( A )$ on ${\mathcal{F}}(L^{2}({\mathbf{R}}))$ such that 
\begin{equation}
\Gamma(e^{i s A }) = e^{i s d\Gamma( A )}   \label{prop2}
\end{equation}
in terms of the self-adjoint operator $A$ that acts on $L^{2}({\mathbf{R}})$.
In particular, it follows that the representation of the
history algebra on the Fock space
${\mathcal{F}}(L^{2}({\mathbf{R}}))$ carries a
(weakly continuous) representation of the one-parameter
family of unitary operators $s\mapsto e^{\frac{i}{\hbar}s
S_{\kappa}} = \Gamma(\Sigma_s)$. Therefore, the generator
${S}_{\kappa}$ also exists on
${\mathcal{F}}(L^{2}({\mathbf{R}}))$, and $S_\kappa =
d\Gamma (-\hbar {\sigma}_{\kappa})$ where
${\sigma}_{\kappa}$ is a self-adjoint operator on
$L^{2}({\mathbf{R}})$ that is defined as
\begin{equation}
 {\sigma}_{\kappa} \psi (t):=
    \left(\omega \kappa (t) - i \frac{d}{dt}\right)\psi(t)
\end{equation}

In what follows, we will restrict our attention to the
particular case $\kappa(t) = 1$ for the simple harmonic
oscillator action operator $S$
 \begin{equation}
   S:= \int^{+\infty} _{-\infty} (p_{t}\dot{x}_{t}-
H_{t})dt  \label{Def:op_S}
 \end{equation}

\section{The Definition of the Liouville Operator}
The first
term of the action operator eq.\ (\ref{Def:op_S}) is
identical to the kinematical part of the classical action
functional eq.\ (\ref{klas}). For reasons that will become apparent later, we write $S_{\kappa}$ as the difference between two
operators: the `Liouville' operator and the Hamiltonian
operator. The Liouville operator is formally written as
\begin{equation}
 V:= \int^{\infty}_{-\infty}(p_{t}\dot{x_{t}}) dt \label{liou}
\end{equation}
where
\begin{equation}
 S_{\kappa} = V - H_{\kappa}.
\end{equation}
We prove the existence of $V$ on
${\mathcal{F}}(L^{2}({\mathbf{R}}))$ using the
same technique as before. Namely, we can see at once that
the history algebra automorphism
\begin{equation}
 e^{\frac{i}{\hbar}sV} b_{f} e^{ - \frac{i}{\hbar}sV} = b_{B_{s}f}
\end{equation}        \label{Vauto}
is unitarily implementable. Here, the unitary operator
$B_s$, $s\in\mathbf{R}$, acting on
$L^{2}({\mathbf{R}})$ is defined by
\begin{equation}
 (B_{s}f)(t):= e^{s\frac{d}{dt}}f(t) = e^{i s D}f(t) = f(t+s)
\end{equation}
where $D:= -i\frac{d}{dt}$. The Liouville operator $V$
has the following commutation relations with the
generators of the history algebra:
\begin{eqnarray}
  [\, V , x_{f}\,] &=& -i\hbar x_{\dot{f}}          \\
 {[}\, V, p_{f}\,] &=& -i\hbar p_{\dot{f}}          \\
 {[}\, V,H_{\kappa}\,] &=& -i\hbar H_{\dot{\kappa}} \\
 {[}\, V, S_{\kappa}] &=& i\hbar H_{\dot{\kappa}}   \\
 {[}\, V, H\,] &=& 0 \\
 {[}\, V, S\,] &=& 0 \\
 {[}\, H, S\,] &=& 0     \label{[H,S]}
\end{eqnarray}
where we have defined $H:= \int^{\infty}_{-\infty}H_{t}dt$. As we shall see, these commutators will play an important role in the physical interpetation of the Liouville operator.

We notice that $V$ transforms, for example, $b_{t}$ from
one time $t$---that refers to the Hilbert space
${\mathcal{H}}_{t}$---to another time $t+s$, that refers to
${\mathcal{H}}_{t+s}$. More precisely, $V$ transforms the
support of the operator-valued distribution $b_{t}$ from
$t$ to $t+s$:
\begin{equation}
 e^{\frac{i}{\hbar}sV} b_{f}
        e^{ - \frac{i}{\hbar}sV} = b_{f_s}
\end{equation}
where $f_s(t):=f(s+t)$. We shall return to the significance
of this later.

\section{The Fourier-transformed n-particle History Propositions}

We shall now briefly consider the eigenvectors of the action operator $S$---as we shall see later, these play a significant part in understanding the role in the quantum history theory of the classical solutions in the equations of motion. An interesting family of history
propositions emerged from the representation space
${\mathcal{F}}[L^{2}({\mathbf{R}} ,dt)]$, acting on
the $\delta$-function normalised basis of states
$|0\rangle$, $|t_{1}\rangle:= b^{\dagger}_{t_{1}}|0\rangle$
, $|t_{1}, t_{2}\rangle:=
b^{\dagger}_{t_{1}}b^{\dagger}_{t_{2}}|0\rangle $ {\em
etc}; or, in smeared form, $|\phi\rangle:=
b^{\dagger}_{\phi}|0 \rangle $ {\em etc}. As noted in Chapter 3, the projection
operator $ |t\rangle \langle t| $ corresponds to the
history proposition `there is a unit energy $\hbar\omega$
concentrated at the time point t'. The physical
interpretation for this family of propositions, was deduced
from the action of the Hamiltonian operator on the family
of $|t\rangle$ states.

To study the behaviour of the $S$ operator, a particularly
useful basis for ${\mathcal{F}}[L^{2}({\mathbf{R}}
, dt)]$ is the Fourier-transforms of the
$|t\rangle$-states. Indeed, if we consider the Fourier
transformations
\begin{eqnarray}
 |\nu \rangle &=& \int^{+\infty}_{-\infty}e^{i\nu
    t}b^{\dagger}_{t}|0\rangle dt                       \\
  |\nu_{1}, \nu_{2}\rangle &=& \int^{+\infty}_{-\infty}
    e^{i\nu_{1}t_{1}} e^{i\nu_{2}t_{2}} b^{\dagger}_{t_{1}}
        b^{\dagger}_{t_{2}}|0\rangle dt_{1}dt_{2}       \\
 b_{\nu}&=& \int^{+\infty}_{-\infty}e^{i\nu t}b_{t} dt  \\
 b^{\dagger}_{\nu}&=& \int^{+\infty}_{-\infty}e^{-i\nu
        t}b^{\dagger}_{t} dt
\end{eqnarray}
the Fourier transformed $|\nu\rangle $- states are defined
by $|\nu\rangle:= b^{\dagger}_{\nu}|0\rangle,
|\nu_{1},\nu_{2}\rangle:=
b^{\dagger}_{\nu_{1}}b^{\dagger}_{\nu_{2}}|0\rangle $ {\em
etc}. These states are eigenvectors of the operator $S$:
 \begin{eqnarray}
   S |0 \rangle &=& 0   \\
   S|\nu\rangle &=& \hbar(\nu-\omega)|\nu\rangle        \\
   S|\nu_{1}, \nu_{2} \rangle &=& \hbar [ (\nu_{1} - \omega) +
    (\nu_{2} - \omega) ] |\nu_{1}, \nu_{2} \rangle      \\
        &\vdots&                                \nonumber
\end{eqnarray}
and we note in particular that
$e^{\frac{i}{\hbar}sS}|0\rangle = |0\rangle$.

The $|\nu \rangle$-states are also eigenstates of the time-averaged history Hamiltonian operator:
\begin{eqnarray}
   H |0 \rangle &=& 0   \\
   H|\nu\rangle &=& \hbar \omega |\nu\rangle    \\
   H|\nu_{1}, \nu_{2} \rangle &=& 2 \hbar
        \omega |\nu_{1}, \nu_{2} \rangle        \\
        &\vdots&                            \nonumber
\end{eqnarray}
which is consistent with these states being eigenstates of $S= V - H$, since $[S,H]=0$ from Eq.\ (\ref{[H,S]}). Again, we see how the
integer-spaced spectrum of the standard quantum theory of the simple harmonic oscillator appears in the HPO theory.

\section{The Velocity Operator}
 The HPO approach to the consistent-histories theory has the striking
feature that, formally, there exists an operator that corresponds to
propositions about the {\em velocity\/} of the system: namely,
$\dot x_t:={d\over dt}x_t$. More rigorously, we can adopt the
procedure familiar from standard quantum field theory and define
\begin{equation} 
	\dot x_f:=-x_{\dot f} \label{Def:dot x}
\end{equation} 
which is meaningful provided that (i) the test-function $f$ is
differentiable; and (ii) $f$ `vanishes at infinity' so that the
implicit integration by parts used in Eq.\ (\ref{Def:dot x}) is
allowed; {\em i.e.}, heuristically, $x_f=\int_{-\infty}^\infty
dt\, x_t f(t)$.

	The rigorous existence of $\dot x_t$ depends on the precise
choice of test-function space used in the smeared form of the HA in
Eqs.\ (\ref{SmearedCtsHWxx})--(\ref{SmearedCtsHWxp}). In the
analogous situation in normal quantum field theory, the
test-functions are chosen so that the spatial derivatives of the
quantum field exist, this being necessary to define the Hamiltonian
operator. In our case, the situation is somewhat different since the
energy operator $H_t$ [see Eq.\ (\ref{Def:Ht})] does not depend on
$\dot x_t$ and hence there is no {\em a priori\/} requirement for
$\dot x_t$ to exist. However, what {\em is\/} clear from Eq.\
(\ref{ctsHWxx}) is that {\em if\/} $\dot x_t$ exists then
\begin{equation}
	{[\,}x_t,\,\dot x_s\,]=0				\label{[xtxdot]} 
\end{equation}
and hence our theory allows for history propositions that include
assertions about the position of the particle and its
velocity at the same time; in particular, the velocity ${\dot x}_t$
and momentum $p_t$ are not related. In this context it should be
emphasised once more that $x_t$, $t\in\mathR$, is a one-parameter
family of {\em Schr\"odinger\/}-picture operators---it is {\em
not\/} a Heisenberg-picture operator, and the equations of motion do
not enter at this level.

There is an interesting approach to understanding Eq.\ (\ref{[xtxdot]}) that comes from the underlying continuous time structure of the theory. As we explained in the previous chapter, because of the history group construction the histories are defined continuously with respect to the time; as a result, we can ask questions about the position $ x_t $ of the system at {\em any\/} time. Hence, using ideas drawn from standard quantum theory, we can produce an operational procedure for defining the velocity $\dot{x}_t$ by using the same {\em measurement apparatus\/} as the one we could have used for evaluating the position of the system.\
In general, the information that we require by asking a question about the system, is determined by the choice of the smearing function for the $x_t$ and $\dot{x}_t$ operators. 

	The existence of a velocity operator that commutes with position
is a striking property of the HPO approach to consistent histories
and raises some intriguing questions. For example, a classic paper
by Park and Margenau \cite{PM68} contains an interesting discussion
of the uncertainty relations, including a claim that it {\em is\/}
possible to measure position and momentum simultaneously provided
the latter is defined using time-of-flight measurements. The
existence in our formalism of the vanishing commutator Eq.\
(\ref{[xtxdot]}) throws some new light on this old discussion. Also
relevant in this respect is Hartle's discussion of the operational
meaning of momentum in a history theory \cite{Har91b}. In
particular, he emphasises that an accurate measurement of momentum
requires a long time-of-flight, whereas---on the other hand---our
definition of velocity as the time-derivative of the history
variable $x_t$ clearly involves a vanishingly small time interval.
Presumably this is the operational difference between momentum and
velocity in the HPO approach to consistent histories.

The existence of the Liouville operator in the
HPO scheme, allows an interesting comparison between the
velocity operator ${\dot{x}}_{f}$ and the momentum operator $p_{f}$: namely, the latter is defined by the
history commutation relation of the position with the
Hamiltonian, while we can define the velocity operator from
the history commutation relation of the position with the
Liouville operator:
\begin{eqnarray}
  [\,x_{f}, H\,] &=& i\hbar \frac{p_{f}}{m}   \\
  {[}\, x_{f}, V \,] &=& i\hbar{\dot{x}}_{f}
\end{eqnarray}
These relations illustrate the different nature of the
momentum $p_{f}$ and the velocity ${\dot{x}}_{f}$: in particular, the  behaviour of the momentum is fundamentally dynamical (as shown by the relation to the Hamiltonian operator), whereas the velocity is fundamentally kinematical (as shown by the relation with the Liouville
operator).

\section{The Heisenberg Picture}

At this point in our discussion, it is useful to investigate the analogue of the Heisenberg
picture in our continuous-time HPO theory. This would help to clarify the relation between momentum and velocity; it will also be a central feature in our discussion in Chapter 6 of the history version of relativistic quantum field theory.

In standard quantum
theory, the Heisenberg-picture version of an operator $A$ is defined
with respect to a time origin $t=0$ as
\begin{equation} 
	A_H(s):=e^{isH/\hbar}\,A\,e^{-isH/\hbar}. \label{Def:AH(t)} 
\end{equation} 
In particular, for the simple harmonic oscillator we have
\begin{eqnarray} 
x(s)&=&\cos[\omega s]x+{1\over m\omega}\sin[\omega s]p \\ 
p(s)&=&-m\omega\sin[\omega s]x+\cos[\omega s]p.
\end{eqnarray}
The Heisenberg-picture operator $x(s)$ satisfies the classical
equation of motion
\begin{equation}
		{d^2x(s)\over ds^2}+\omega^2 x(s)=0,		\label{EM} 
\end{equation}
and the commutator of these operators is 
\begin{equation}
	{[\,}x(s_1),\,x(s_2)\,]={i\hbar\over m\omega}\sin[\omega(s_1-s_2)]
							\label{[xs1xs2]}
\end{equation}
which, on using the Heisenberg-picture equation of motion
\begin{equation}
	p:=m\left.{dx(s)\over ds}\right|_{s=0},		\label{Def:p} 
\end{equation}
reproduces the familiar canonical commutation relation Eq.\
(\ref{CCR}).

	In trying to repeat this construction for the history theory we
might be tempted to define the Heisenberg-picture analogue of, say,
$x_t$ as
\begin{equation}
	x_{H,t}(s):= e^{isH_t/\hbar}\,x_t\, e^{-isH_t/\hbar}.
\end{equation}
However, this expression is not well-defined since it corresponds to
choosing the test-function in Eq.\ (\ref{UxtU}) as
$\chi(t'):=s\delta(t-t')$, which leads to ill-defined products of
$\delta(t-t')$.

	What is naturally suggested instead is to define `time-averaged'
Heisenberg quantities
\begin{equation}
	x_{\kappa,t}:=e^{iH_\kappa/\hbar}\,x_t\,e^{-iH_\kappa/\hbar}
		=\cos[\omega\kappa(t)]x_t+
		{1\over m\omega}\sin[\omega\kappa(t)]p_t \label{Def:xkt}
\end{equation}
for suitable test functions $\kappa$. The analogue of the equation
of motion Eq.\ (\ref{EM}) is the functional differential
equation
\begin{equation} 
{\delta^2 x_{\kappa,t}\over\delta\kappa(s_1)\delta\kappa(s_2)} +
\delta(t-s_1)\delta(t-s_2)\omega^2x_{\kappa,t}=0,
\end{equation}
while the history analogue of Eq.\ (\ref{Def:p}) is
\begin{equation}
	\delta(t-s)p_t=m \left.{\delta
				x_{\kappa,t}\over\delta\kappa(s)}\right|_{\kappa=0}, 
\end{equation}
and the analogue of the `covariant commutator' Eq.\
(\ref{[xs1xs2]}) is
\begin{equation}
{[\,}x_{\kappa_1,t_1},\,x_{\kappa_2,t_2}\,]={i\hbar\over m\omega}
\delta(t_1-t_2)\sin[\omega(\kappa_1(t_1)-\kappa_2(t_2)]
\end{equation} 
which correctly reproduces the canonical history algebra.
  However, the use of the expression Eq.\ (\ref{Def:xkt}) to define a `Heisenberg-picture' operator, lacks the analogy with the classical equations of motion Eq.\ (\ref{EM}).
  
Before developing this point further however, it is worth noting that {\em any\/} definition of a Heisenberg-picture in the HPO theory will involve two time labels: an `external' label $t$---that specifies the time the proposition is asserted---and an `internal' label $s$ that, for a fixed time $t$, is the time parameter of the Heisenberg picture associated with the copy ${\mathcal{H}}_{t}$ of the standard Hilbert space. Using the results from the history algebra automorphisms Eqs.\ (\ref{UbtU}, \ref{Vauto}, \ref{Sauto}) for the definition of the $H$, $V$ and $S$ operators, it can be seen that the two labels appear naturally in the final version of the Heisenberg picture: they are related to
the groups that produce the two types of time
transformations. In addition, the analogy with the
classical expressions is regained.

To see this explicitly, we define a Heisenberg-picture
analogue of $x_{t}$ as
\begin{eqnarray}
x_{\kappa,t,s } :&=& e^{\frac{i}{\hbar} s H_{\kappa}} x_{t}
e^{-\frac{i}{\hbar}sH_{\kappa}} \\  \label{heis}
&=& \cos[\omega s \kappa(t)]x_{t} + \frac{1}{m\omega}\sin[\omega s \kappa(t)]p_{t}   \nonumber  \\
p_{ \kappa,t,s } :&=& e^{ \frac{i}{\hbar} s H_{\kappa}}
p_{t} e^{-\frac{i}{\hbar} s H_{\kappa}} \\
    &=& -m \omega \sin [\omega s \kappa(t)]x_{t} + \cos [\omega s \kappa(t)]p_{t} \nonumber
\end{eqnarray}
where $\kappa(t)$ is now a `fixed' function.
The commutation relations for these operators are
\begin{eqnarray}
 [\,x_{\kappa,t}(s),x_{\kappa^{\prime},t^{\prime}}(s^{\prime})\,]
        &=& \frac{i\hbar}{m\omega}\sin[\omega\kappa
            (s^{\prime}-s)]\delta(t-t^{\prime})             \\
    {[}\,x_{\kappa,t}(s), S_{{\kappa}^{\prime}}\,] &=&
     i\hbar \Big{[} \cos[s\omega \kappa(t)] \dot{x_{t}}
      + \frac{1}{m\omega} \sin[s\omega \kappa(t)]
        \dot{p_{t}} - \frac{{\kappa}^{\prime}}{m}
            p_{\kappa , t, s} \Big{]}  \hspace{1cm}                     \\
    {[}\,p_{\kappa,t}(s), S_{{\kappa}^{\prime}}\,] &=&
        i\hbar\Big{[}\cos[s\omega \kappa(t)] \dot{p_{t}} -
        m\omega \sin[s\omega \kappa(t)] \dot{x_{t}} +
        {\kappa}^{\prime}(t) x_{\kappa,t,s}\Big{]}\hspace{1cm}
\end{eqnarray}

which are clearly compatible with  the HPO analogue of
the equations of motion
\begin{equation}
 \frac{d^{2}}{ds^{2}}x_{\kappa, t, s}+ \omega^{2} {\kappa(t)}^{2} x_{\kappa, t, s}= 0
\end{equation}
which follow directly from the definition of the operator $x_{\kappa, t, s}$ in Eq.\ (\ref{heis}).
We notice the strong resemblance with standard quantum
theory; for the case $\kappa(t)=1$, the classical
expressions are fully recovered.

\chapter{Time Transformations in the HPO}

\section{Introduction}

One exciting feature of the HPO theory is the way
that the time transformations appear in the
formalism.

In what follows, we will show that, constructed as
a quantum analogue of the classical action
functional, $S_{\kappa}$ does indeed act as a
generator of time-transformations in the HPO
theory. Furthermore, we will argue that
$S_{\kappa}$ is related to the two laws of
time-evolution in standard quantum theory namely,
state-vector reduction, and unitary time-evolution
between measurements.

A comparison with the classical theory seems
appropriate at this point; thus, we present a
classical analogue of the HPO where the
continuous-time classical histories can be seen as
analogues of the continuous-time quantum
histories.

We further exploit the above analogy to discuss
the `classical' behaviour of the history quantum
scheme. In particular, we expect the action
operator to be involved in some way with the
dynamics of the theory. To this end, we show how
the action operator appears in the expression for
the decoherence functional, with operators acting
on coherent states, as used by Isham and Linden
\cite{IL95}.

\section{The Two Types of Time Transformation}

In standard classical mechanics, the Hamiltonian
$H$ is the generator of time transformations. In
terms of Poisson brackets, the generalised
equation of motion for an arbitrary function $u$
is given by
\begin{equation}
 \frac{du}{dt} = \{ u , H \} + \frac{\partial u}{\partial t}.
\end{equation}

In a HPO theory, the Hamiltonian operator $H_t$
produces phase changes in time, preserving the
time label $t$ of the Hilbert space on which, at
least formally, $H_t$ is defined. On the other it
is the Liouville operator $V$ that assigns,
analogous to the classical case, history
commutation relations, and produces time
transformations `from one Hilbert space to
another'. The action operator generates a
combination of these two types of
time-transformation. If we use the notation
$x_{f}(s)$ for the $\it{history}$
Heisenberg-picture operators smeared with respect
to the time label $t$ Eq.\ (\ref{heis}) (with
$\kappa=1$), we observe that they behave as
standard Heisenberg-picture operators, with time
parameter $s$. A novel result however is the
observation that, the operator $x_{f}(s)$ for
instance, changes with respect to time parameter
$t$ also, in the sense that at a later time
$t^{\prime}= t + \a $ the operator valued
distribution $x_t$ is smeared by the function
$f^{\prime}(t^{\prime})= f(t + \a)$. Furthermore,
their history commutation relations are
\begin{eqnarray}
[\,x_{f}(s) , V\,] &=& i\hbar {\dot{x}}_{f}(s) \\
  {[}\, x_{f}(s), H\,] &=& \frac{i\hbar}{m}p_{f}(s)         \\
  {[}\, x_{f}(s), S\,] &=&
    i\hbar({\dot{x}}_{f}(s)- \frac{1}{m}p_{f}(s))
\end{eqnarray}
which, as we shall see, strongly resemble the
corresponding expressions in the classical history
theory which we shall develop.

We now define a one-parameter group of
transformations $T_{V}(\tau)$, with elements
$e^{\frac{i}{\hbar}\tau V}$, $\tau\in{\mathbf{R}}$
where $V$ is the Liouville operator Eq.\
(\ref{liou}), and we consider its action on the
Heisenberg-picture operator $b_{t,s}$; for
simplicity we write the unsmeared expressions
\begin{equation}
 e^{\frac{i}{\hbar}\tau V} b_{t,s} e^{ - \frac{i}{\hbar}\tau V} = b_{t+\tau,s},
\end{equation}
which makes particularly clear the sense in which
the Liouville operator is the generator of
transformations of the time parameter $t$
labelling the Hilbert spaces ${\mathcal{H}}_{t}$.

Next we define a one-parameter group of
transformations $T_{H}(\tau)$, with elements
$e^{\frac{i}{\hbar}\tau H}$, where $H$ is the
time-averaged Hamiltonian operator
\begin{equation}
 e^{\frac{i}{\hbar}\tau H} b_{t,s} e^{ - \frac{i}{\hbar}\tau H} = b_{t,s+\tau}.
\end{equation}
Thus the Hamiltonian operator is the generator of
phase changes of the time parameter $s$, produced
only on one Hilbert space ${\mathcal{H}}_{t}$, for
a fixed value of the `external' time parameter
$t$.

 Finally, we define the one-parameter group of
transformations $T_{S}(\tau)$, with elements
$e^{\frac{i}{\hbar}\tau S}$, where $S$ is the
action operator, which acts as
\begin{equation}
 e^{\frac{i}{\hbar}\tau S} b_{t,s} e^{ - \frac{i}{\hbar}\tau S} = b_{t+\tau , s+\tau}
\end{equation}
We see that the action operator generates both
types of time transformations---a feature that
appears only in the HPO scheme.

In Fig.1a,b, we represent the tensor product of
Hilbert spaces as a sequence of planes (each one
representing a copy of the standard Hilbert
planes), and a quantum continuous-time history as
a curve in that space. Each plane is labeled by
the time label $t$ that the corresponding Hilbert
space $H_t$ carries. Thus a history is depicted as
a curve along an $n$-fold sequence of `Hilbert
planes' ${\mathcal{H}}_{t_{i}}$. As we will
explain later, in analogy to this, we can
represent a classical history as a curve along an
$n$-fold sequence of planes corresponding to
copies of the standard phase-space
$\Gamma_{t_{i}}$, as we will explain later. The
time transformations generated by the Liouville
operator, shift the path in the direction of the
`Hilbert planes'. On the other hand, the
Hamiltonian operator generates time
transformations that move the history curve in the
direction of the path, as represented on one
`Hilbert plane'.

\begin{figure}[h!]
\centerline{\psfig{file=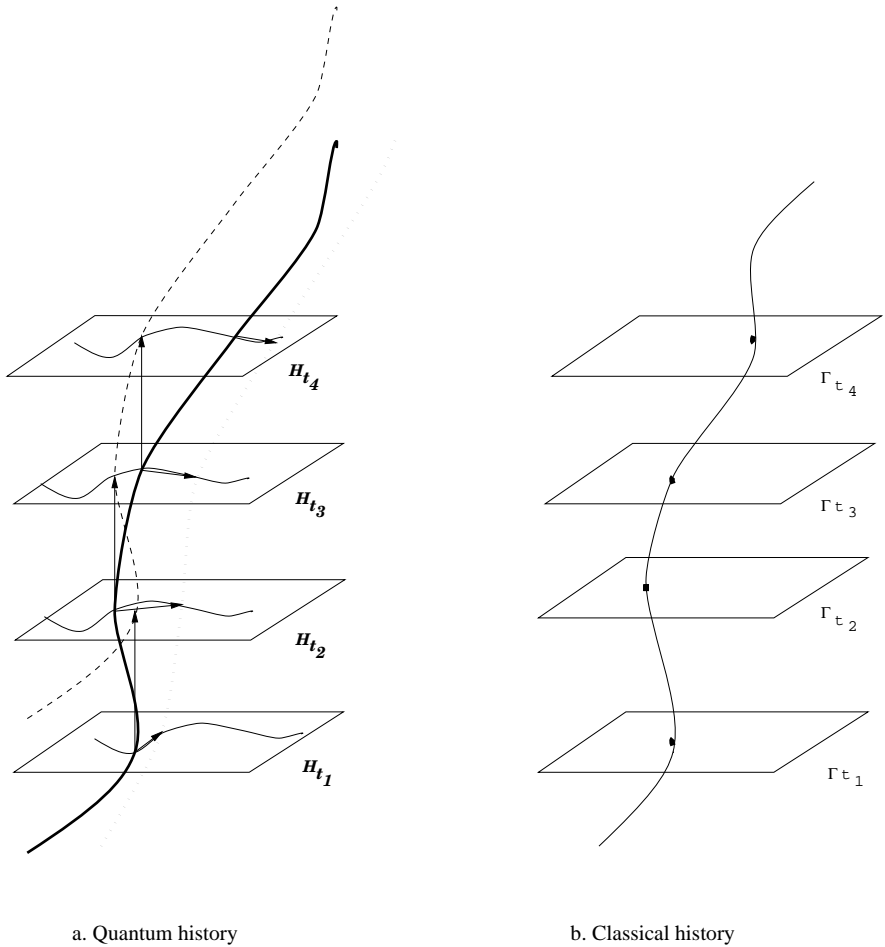,width=4.6in}}
\caption{Quantum
and classical history curves. In Figure 1.a the transformation
of the history curves generated by $V$ is represented by the
dashed line, while the transformation generated by $H$ are
represented by the dotted line. The curves drawn on each `Hilbert plane' correspond to the Hamiltonian transformations as effected on
the corresponding Hilbert space. In Figure 1.b the classical history remains invariant under the corresponding time transformations }
\label{fig:test}
\end{figure}

\subsection{The Two-fold Time Interpretation}

In standard quantum theory, time-evolution is
described by two different laws: the state-vector
reduction that occurs when a measurement is made,
and the unitary time-evolution that takes place
between measurements. Thus, according to von
Neumann, one has to augment the Schr\"odinger
equation with a collapse of the state vector
associated with a measurement \cite{9}.

I would like to claim that the two types of
time-transformations observed in the HPO theory
are associated in some way with the two dynamical
processes in standard quantum theory: the time
transformations generated by the Liouville
operator $V$ are related to the state-vector
reduction (more precisely, the time ordering
implied by the state-vector reduction), while the
time transformations produced by the Hamiltonian
operator $H$ are related to the unitary
time-evolution between measurements.

The argument in support of this assertion is as
follows. Keeping in mind the description of the
history space as a tensor product of single-time
Hilbert spaces ${\mathcal{H}}_{t}$, the $V$
operator acts on the Schr\"odinger-picture
projection operators by translating them in time
from one Hilbert space to another. These
time-ordered projectors appear in the expression
for the decoherence functional that defines
probabilities. In history theory, the expression
for probabilities in a consistent set, is the same
as that derived in standard quantum theory using
the projection postulate on a time-ordered
sequence of measurements \cite{Har93a,GH90b}. It
is this that suggests a relation of the Liouville
operator to `state-vector reduction'.

Indeed, the class operator in Eq.\
(\ref{Def:C_a})that represents a history as a
time-ordered sequence of Schr\"odinger-picture
operators interleaved with the unitary-time
operator, was constructed in a way that imitates
the state-vector reduction (time evolution) of the
standard quantum theory. In the HPO theory, for a
set of consistent histories, the Liouville
operator is the generator of time transformations
that takes a Schr\"odinger-picture operator at
some time $t$ (corresponding to the Hilbert space
copy with the same t-time label, ${\mathcal
H}_{t}$ ), to another time $t^{\prime}$
(corresponding to the Hilbert space copy with the
same t-time label ${\mathcal{H}}_{t^{\prime}}$).
Following the description for the class operator
in generalised consistent histories, we argue that
in a similar way, the action of the Liouville
operator can be related to the proccess of
state-vector reduction.

To strengthen this claim, in what follows we will
show the analogy of $V$ with the $S_{cts}$
operator (an approximation of the derivative
operator that we shall define shortly) that
appears in the decoherence function in the HPO
formalism, and is implicitly related to the
state-vector reduction by specifying the
time-ordering of the action of the single-time
projectors. The action of $V$ as a generator of
time translations depends on the partial (in fact,
total) ordering of the time parameter treated as
the causal structure in the underlying spacetime.
Hence, the $V$-time translations illustrate the
purely kinematical function of the Liouville
operator.

The Hamiltonian operator producing
transformations, via a type of Heisenberg
time-evolution, appears as the `clock' of the
theory. As such, it depends on the particular
physical system that the Hamiltonian describes.
Indeed, we would expect the definition of a
`clock' for the evolution in time of a physical
system to be connected with the dynamics of the
system concerned. We note that the smearing
function $\kappa (t)$ used in the definition of
the Hamiltonian operator can be interpreted as a
mechanism of implementing the idea of idea of
reparametrizing time---a concept that plays a key
role in quantum gravity; in the present context
however, $\kappa$ is kept fixed for a particular
physical system.

The coexistence of the two types of
time-evolution, as reflected in the action
operator identified as the generator of such time
transformations, is a striking result. In
particular, its definition is in accord with its
classical analogue, namely the Hamilton action
functional. In classical theory, a distinction
between a kinematical and a dynamical part of the
action functional also arises in the sense that
the first part corresponds to the symplectic
structure and the second to the Hamiltonian.

\section{The Classical Imprint of the HPO}
Let us now consider more closely the relation of
the classical and the quantum histories. We have
shown above how the action operator generates time
translations from one Hilbert space to another,
through the Liouville operator; and on each
labeled Hilbert space ${\cal{H}}_{t}$, through the
Hamiltonian operator. We now wish to discuss in
more detail the analogue of these transformations
in the classical case.

We recall that a history is a time-ordered
sequence of propositions about the system. The
continuous-time quantum history in the HPO system,
makes assertions about the values of the position
or the momentum of the system, or a linear
combination of them, at each moment of time, and
is represented by a projection operator on the
continuous tensor product of copies of the
standard Hilbert space.

This raises the intruiging question of the extent
to which one expects a continuous-time classical
history theory should reflect the underlying
temporal logic of the situation. Thus the
assertions about the position and the momentum of
the system at each moment of time should be
represented on an analogous history space: as we
shall see, this can be achieved by using the
Cartesian product of a continuous family (labelled
by the time $t$) of copies of the standard
classical state space.

In classical mechanics, a (fine-grained) classical
history is represented by a path in the state
space. Indeed, a path $\gamma$ is defined as a map
from the real line into the standard phase-space
$\Gamma$:
\begin{eqnarray}
     \gamma : & {\mathR}& \to \Gamma    \\
     & t& \mapsto (q(\gamma(t)), p(\gamma(t)))    \nonumber
\end{eqnarray}
where $q^i((\gamma(t))$ and $p_j(\gamma(t))$,
$i,j=1,2,\ldots,N$ (where the dimension of
$\Gamma$ is $2N$) are the position and momentum
coordinates\footnote{The notation here is somewhat
cryptic but hopefully the intention will be
apparent.} of the path $\gamma$, at the time $t$.
For our purposes, we shall consider the path
$t\mapsto\gamma(t)$ to be defined for $t$ in some
finite time interval $[t_1, t_2]$. We shall denote
by $\Pi$ the set of all such $C^\infty$ paths.

The key idea of this new approach to classical
histories is contained in the symplectic structure
of the theory: the choice of the Poisson bracket
must be such that it includes entries at different
moments of time. Thus we suppose that the space of
functions on $\Pi$ is equipped with the `history
Poisson bracket' defined by
\begin{equation}
  \mbox{\boldmath$\{$ } q^i_t , p_{j,t^{\prime}}\mbox{\boldmath$\}$ } =\delta^i_j \delta(t-t^{\prime})
\end{equation}
where we defined the functions $q_t$ on $\Pi$ as
\begin{eqnarray*}
            q_t^i : &\Pi& \to {\mathR} \\
                  &\gamma& \mapsto q^i_t(\gamma):=
q(\gamma(t))^i
\end{eqnarray*}
and similarly for $p_t$.

\paragraph*{The Temporal Logic and the HPO Classical Histories Proposal}
Before proceeding any further with this
construction, it is worth mentioning that the
temporal logic of classical history theory can be
defined without any reference to the quantum case.

We start by recalling that there exists a
correspondance between single-time history
propositions $P$ and subsets\footnote{In a more
precise treatment the restriction is often imposed
that the subsets concerned should be {\em Borel\/}
subsets.} of the phase space $\Gamma$: a
proposition is represented by a characteristic
function $ \chi_{C}(s) $, defined as \beq
    \chi_{C}(s)\eqdef\left\{ \begin{array}{ll}
             1 \quad \mbox{if $s\in C$},        \\
             0 \quad \mbox{if $s\notin C$}.
             \end{array}
\right. \eeq where $C$ is a subset of the state
space $\Gamma$.

Let us suppose that the proposition $P$
corresponds to the subset $C_{P}\subset \Gamma$,
and the proposition $Q$ corresponds to the subset
$C_{Q}\subset \Gamma$. Then, it is obvious that
the proposition ``P and Q" corresponds to the
subset $C_{P}\cap C_{Q}\subset \Gamma$.

Next, we aim to define temporal propositions: for
instance ``P at time $t_1$ and then Q at time
$t_2$". For a state $ s\in \Gamma $, the
proposition is true if $s\in C_P$ and $ s\in
\tau_{\{t_1, t_2\}}(C_Q) $, where
 \beqa
    \tau_{\{t,t^{\prime}\}}: &\Gamma& \mapsto \Gamma  \nonumber  \\
     &s_t & \leadsto \tau_{\{t,t^{\prime}\}}(s_t) := s_{t^{\prime}}.
 \eeqa
denotes the time-development map that evolves a
state $s\in \Gamma$ at time $t$, to the
corresponding state $\tau_{\{t,t^{\prime}\}}(s)$
at time $t^{\prime}$.

Hence, from this perspective the temporal
proposition ``P at time $t_1$ and then Q at time
$t_2$" corresponds to the subset \beq
    C_P \cap \tau^{-1}_{\{t_1, t_2\}} (C_Q)
\eeq This can be regarded as a classical analogue
of a quantum class operator.

Furthermore, let us suppose that, the proposition
$Q$ is in fact, the proposition $\tau_{\{t_1,
t_2\}}(P)$ to be true at time $t_2$. Then the
corresponding subset of the temporal proposition
``P at time $t_1$ and then Q at time $t_2$" is
\beq
 C_P \cap \tau^{-1}_{\{t_1, t_2\}} (\tau_{\{t_1, t_2\}}(C_P)) = C_P \cap C_P = C_P
\eeq Hence, the problem that arises is how to
discriminate the above temporal proposition from
the proposition ``P is true at time $t_1$". From
another perspective, what we seek is another
representation of temporal logic that is
independent of the specific dynamics of a
particular system.

Actually, the natural solution to this, is to take
the Cartesian product of copies of the phase
space. More precisely, we employ a mathematical
model in which each single-time proposition
corresponds to a subset of a {\em particular
copy\/} of the phase space $\Gamma_t$, labeled by
the time parameter. The proposition ``$P_{t_1}$
and then $P_{t_2}$ and then...and then $P_{t_n}$"
corresponds to the subset \beq C_{t_1}\times
C_{t_2}\times\cdots\times\cdots C_{t_n}
 \eeq of
the Cartesian product
$\Gamma_{t_1}\times\Gamma_{t_2}\times\cdots\times\Gamma_{t_n}$
of copies of the standard phase space.

We know that a natural way to proceed with the
quantisation algorithm for such formalism, is to
take the tensor product of copies of Hilbert
spaces ${\mathcal{H}}_{t_1}\otimes
{\mathcal{H}}_{t_2}\otimes\cdots\otimes
{\mathcal{H}}_{t_n}$. This is an important result
because it provides another, rather novel
justification of the original quantum construction
of the HPO theory. Furthermore it shows that, in
general, one may well start by defining classical
HPO histories, and then proceed with their quantum
analogues.

\paragraph*{The Classical Analogue of the HPO Theory}
 We now define the history action functional $S_{h}(\gamma)$ on
 $\Pi$ as
\begin{equation}
   S_{h}(\gamma):=
   \int^{t_2}_{t_1}{[p_t\dot{q_t}-H_t(p_t,q_t)](\gamma)\,dt}
\end{equation}
where $q_t(\gamma)$ \footnote{This is a compact
expression, where the indices $i$,$j$ used
previously are implicit.}is the position
coordinate at the time point $t \in [t_1, t_2]$ of
the path $\gamma$, and $\dot{q_t}(\gamma)$ is the
velocity coordinate at the time point $t\in[t_1,
t_2]$ of the path $\gamma$.

We also define the history classical analogues for
the Liouville and time-averaged Hamiltonian
operators as
\begin{eqnarray}
 V_{h} ( \gamma )&:=& \int^{t_2}_{t_1}{[p_t\dot{q_t}](\gamma)\,dt}    \\
 H_{h} ( \gamma )&:=& \int^{t_2}_{t_1}{[H_t(p_t,q_t)](\gamma)\,dt}    \\
   S_{h}( \gamma ) &=& V_{h}( \gamma )- H_{h}( \gamma )
\end{eqnarray}

In classical mechanics, the least action principle
states that there exists a functional $S(\gamma )=
\int^{t_2}_{t_1}{[p\dot{q}-H(p,q)](\gamma)\,dt}$
such that the physically realised path is a curve
in state space, ${\gamma}_{0}$, with respect to
which the condition $\delta S ({\gamma}_{0}) = 0$
holds, when we consider variations around this
curve. From this, the Hamilton equations of motion
are deduced to be
\begin{eqnarray}
  \dot{q} &=& \{ q , H \}         \\
  \dot{p} &=& \{ p , H \}
\end{eqnarray}
where $q$ and $p$---the coordinates of the
realised path ${\gamma}_{0}$---are the solutions
of the classical equations of motion. For any
function $F(q,p)$ of the classical solutions it is
also true that
\begin{equation}
    \{F, H \} = \dot{F}
\end{equation}

In the case of classical continuous-time
histories, one can formulate the above variational
principal in terms of the Hamilton equations with
the statement: A classical history ${\gamma}_{cl}$
is the realised path of the system---{\em i.e.\/}
a solution of the equations of motion of the
system---if it satisfies the equations
\begin{eqnarray}
     \{q_t , V_{h}\}(\gamma_{cl}) = \{q_t , H_{h}\}(\gamma_{cl}) \label{Ham1}    \\
      \{p_t , V_{h}\}(\gamma_{cl}) = \{p_t , H_{h}\}(\gamma_{cl}) \label{Ham2}
\end{eqnarray}
where ${\gamma}_{cl}= t\mapsto(q_t({\gamma}_{cl})
, p_t({\gamma}_{cl}) )$, and $q_t({\gamma}_{cl})$
is the position coordinate of the realised path
${\gamma}_{cl}$ at the time point $t$. The eqs.\
(\ref{Ham1}--\ref{Ham2}) are the history
equivalent of the Hamilton equations of motion.
Indeed, for the case of the simple harmonic
oscillator in one dimension the eqs.\
(\ref{Ham1}--\ref{Ham2}) become
\begin{eqnarray}
     {\dot{q}}_t ({\gamma}_{cl}) &=& \frac{p_t}{m}({\gamma}_{cl})   \\
     {\dot{p}}_t ({\gamma}_{cl}) &=& - m {\omega}^{2} q_t ({\gamma}_{cl})
\end{eqnarray}
where ${\dot{q}}_t ({\gamma}_{cl}) =
\dot{q}({\gamma}_{cl}(t))$ is the value of the
velocity of the system at time $t$. One would have
expected the result in eqs.\
(\ref{Ham1}--\ref{Ham2}) for the classical
analogue of the histories formalism, as it shows
that the classical analogue of the two types of
time-transformation in the quantum theory
coincide.

 From the eqs.\
(\ref{Ham1}--\ref{Ham2}) we also conclude that the
canonical transformation generated by the history
action functional $S_{h}({\gamma}_{cl})$, leaves
invariant the paths that are classical solutions
of the system:
\begin{eqnarray}
     \{q_t , S_{h} \}(\gamma_{cl})&=& 0  \label{clsol1} \\
     \{p_t , S_{h} \}(\gamma_{cl})&=& 0  \label{clsol2}
\end{eqnarray}
It also holds that any function $F$ on $\Pi$
satisfies the equation
\begin{equation}
     \{ F, S_{h}\} (\gamma_{cl})= 0
\end{equation}

Some of these statements are implicit in previous
work by C. Anastopoulos \cite{6}; an interesting
application of a similar extended Poisson bracket
using a different formulation has been done by I.
Kouletsis \cite{7}.

\subsection{Classical Coherent States for the Simple Harmonic Oscillator}

In the case of the simple harmonic oscillator, the
relation between the classical and the quantum
history theories can be further exemplified by
using coherent states. This special class of
states was used in \cite{IL95} to represent
certain continuous-time history propositions in
the history space. Coherent states are
particularly useful for this purpose since they
form a natural (over-complete) base for the Fock
space representation of the history algebra.

A class of coherent states in the relevant Fock
space is generated by unitary transformations on
the cyclic vacuum state:
\begin{equation}
          |f,h\rangle := U[f,h]|0\rangle
\end{equation}
where $ U [f,h]$ is the Weyl operator defined as
 \begin{equation}
    U [f,h]:= e^{\frac{i}{\hbar}(x_f- p_h)},
  \end{equation}
where $f$ and $h$ are test functions in
$L^2({\mathbf{R}})$. The Weyl generator
 \begin{equation}
      \alpha (f,h):= x (f)- p (h)
 \end{equation}
 can alternatively be written as
 \begin{equation}
 \alpha (f,h)= \frac{\hbar}{i}(b^{\dagger}(w)-b(w^{*}))
 \end{equation}
where $w:=f+ih$.

 Suppose now that, for a pair of functions $(f,h)$, the
operator $\alpha(f,h)$ commutes with the action
operator $S$:
\begin{equation}
   [\,S, \alpha(f,h)\,]= 0.    \label{alpha(f,h)}
\end{equation}
Then any pair $(f,h)$ satisfying this equation is
necessarily a solution of the system of
differential equations:
\begin{eqnarray}
              \dot{f}+ m\omega^2h &=& 0 \label{system1} \\
              \dot{h}- \frac{f}{m} &=& 0   \label{system2}
\end{eqnarray}
We see that if we identify $f$ with the classical
momentum $p_{cl}$ and $h$ with the classical
position $x_{cl}$, then the eqs.\ (\ref{system1}--
\ref{system2}) are precisely the classical
equations of motion for the simple harmonic
oscillator:
\begin{equation}
 \ddot{x_{cl}} + \omega^2 x_{cl}= 0.
\end{equation}

The classical solutions $(f,h)$ distinguish a
special class of Weyl operators ${\alpha}_{cl}
(f,h)$, and hence a special class of coherent
states:
\begin{equation}
  | \exp z_{cl} \rangle:= U_{{\alpha}_{cl} (f,h)}|0 \rangle
\end{equation}
where $z_{{\rm cl}}:=f+ih$.

These classical-like features stem from the
following relation with $S$
\begin{equation}
  [\,S , U_{{\alpha}_{cl}}\,] = 0
\end{equation}
\begin{equation}
  [\,S , P_{|\exp z _{cl}\rangle}\,] = 0  \label{acpr}
\end{equation}
where $ P_{|\exp z_{cl} \rangle}$ is the
projection operator onto the (non-normalised)
coherent state $|\exp z_{cl}\rangle $:
\begin{equation}
   P_{|\exp z_{cl}\rangle}:= \frac{|\exp z_{cl}\rangle
   \langle \exp z_{cl} |}{\langle \exp z_{cl}|\exp
   z_{cl}\rangle}   \label{proj}
\end{equation}

We note that there exists an analogy between eqs.\
(\ref{clsol1}--\ref{clsol2}) and eq.\
(\ref{acpr}), if we consider $(f,h)$ to be the
classical solution: $t\mapsto (q_t,
p_t)(\gamma_{cl})$. In classical histories, the
canonical transformation eqs.\ (\ref{clsol1}--
\ref{clsol2}) generated by the history action
functional vanishes on a solution to the equations
of motion. On the other hand, when we deal with
quantum histories, the action operator produces
the classical equations of motion eqs.\
(\ref{system1}--\ref{system2}) when we require
that it commutes with the projector (as in \ eq.\
(\ref{proj})) which corresponds to a classical
solution $(f,h)$ of the system. However, this is
not directly related to the actual classical limit
of the theory: to make any such physical
predictions we must involve the decoherence
functional and the coarse graining operation.

Notice that many facets of the construction above
hold for a generic potential, as long as there
exists a representation on $ {\cal{V}}_{cts}$ of
the history algebra on which the action operator
is defined. However, it is a subject for future
research to uncover the analogue of the
coherent-states construction in such situations.

\section{The Decoherence Function Argument}

In the consistent histories quantum theory, the
dynamics of a system is described by the
decoherence functional. In a classical theory it
is the action functional that plays a similar role
in regard to the dynamics of the system. It is
only natural then, to seek for the appearance of
the action operator in the decoherence functional.
The aim in this section is to write the HPO
expression for the decoherence functional with
respect to an operator that includes $S$, and to
compare this operator ({\em i.e.}, its matrix
elements), with the operator ${\mathcal{S}}_{cts}
\mathcal{U}$ that appears in the decoherence
functional \cite{IL95}. Hence, it is useful to
present beforehand the definition of the history
propositions and of the decoherence functional for
the special case of the coherent-states, as
presented in \cite{IL95}.

\subsection{The Coherent States History Propositions}
In the construction of the history space $\Vcts =
\otimes_{t\in\mathR} {\mathcal{H}}_t $ in section
3.3, the use of the coherent states played an
important role in demonstrating the isomorphism
  \beqa
    \Vcts\eqdef\ot_t\lbra L^2_t(\mathR)\rbra
        & \simeq & \exp\lbra L^2(\mathR,dt)\rbra. \label{isoh} \\
        \otimes_t \ket{\exp\phi_t} & \mapsto & \ket{\exp\phi(\cdot)} \nonumber.
\eeqa The (non normalised) exponential state
$\ket{\exp\phi_t} \in L^2_t(\mathR)$ can be
written in terms of the normalised coherent states
$\ket{\m(t)}$ as \beq
 \ket{\exp\m(t)} = e^{\frac{1}{2}|\m|^2} \ket{\m(t)}
\eeq The usual normalised coherent states
$\ket{\m(t)}$ are defined as \beq
 \ket{\m(t)} =  e^{\frac{1}{2}|\m|^2 + \m(t)a^{\dagger}}\ket{0}
\eeq with inner product \beq
 \braket{\l(t)}{\m(t)} = e^{\l(t)^* \m(t) - \frac{1}{2}|\l(t)|^{2} - \frac{1}{2}|\m(t)|^{2}}
\eeq The isomorphism Eq.\ (\ref{isoh}) allows one
to identify the projector $P_{\otimes_t
\ket{\exp\l(t)}}$ onto the vector $\otimes_t
\ket{\exp\l(t)}$ in $\otimes_t L^2_t(\mathR)$,
with the projector $P_{\ket{\exp\l(\cdot)}}:=
e^{-\langle\l,\l\rangle}
\ket{\exp\l(t)}\bra{\exp\l(t)}$ onto the vector
$\ket{\exp\l(\cdot)}$ in $\exp{ L^2_t(\mathR ,
dt)}$. The action of the latter is \beq
P_{\ket{\exp\l(\cdot)}}\ket{\exp\m(\cdot)} =
        e^{\inner{\l}{\m-\l}}\ket{\exp\l(\cdot)}
\eeq Furthermore, the projector
$P^{[a,b]}_{\ket{\exp\l(\cdot)}}$, corresponding
to history propositions that involve a {\em
finite\/} time integral $[a,b]$ was defined in
such a way that it is ``active'' in the region
$[a,b]$, otherwise it is equal to the unit
operator \beq
    P^{[a,b]}_{\ket{\exp\l(\cdot)}} \ket{\exp\m(\cdot)}\eqdef
        \exp\left(\int_a^b \l^*(t)(\m(t)-\l(t))\,dt\right)
            \ket{\exp\l\star\m(\cdot)}                                      \label{Pabl}
\eeq where \beq
    (\l\star\m)(t)\eqdef\left\{ \begin{array}{ll}
             \l (t)\quad \mbox{if $t\in [a,b]$},        \\
             \m (t)\quad \mbox{otherwise}.
             \end{array}
\right. \eeq

\subsection{The Coherent States Decoherence Functional}
As we have discussed earlier, in the Gell-Mann and
Hartle generalised histories \cite{GH90b, Har93a},
the decoherence functional$d_{(H,\rho)} (\a,\b)$
is a complex-valued function of pairs of
homogeneous histories $(\hh \a n)$ and
$(\b_{t_1^\prime},\b_{t_2^\prime},\ldots,\b_{t_{m}^\prime})$,
defined as
\begin{equation}
    d(\alpha,\beta)= {\rm tr}(\tilde C_\alpha^\dagger\rho
        \tilde C_\beta)                             \label{Def:d}
\end{equation}
where $\rho$ is the initial density-matrix, and
where the class operator $\tilde C_\alpha$ is
defined in terms of the standard
Schr\"odinger-picture projection operators
$\alpha_{t_i}$ as
\begin{equation}
    \tilde C_\alpha:=U(t_0,t_1)\alpha_{t_1} U(t_1,t_2)
    \alpha_{t_2}\ldots U(t_{n-1},t_n)\alpha_{t_n}U(t_n,t_0),
                                                    \label{Def:C_a}
\end{equation}
where $U(t,t')=e^{-i(t-t')H/\hbar}$ is the unitary
time-evolution operator from time $t$ to $t'$.

As shown in section 2.3, for the case of discrete
time histories on the discrete tensor product of
the standard Hilbert space $\V$ \cite{ILS94}, the
decoherence functional can be written as \beq
    d(\a,\b) = \tr_{(\ot^n \H)\ot(\ot^m \H)}\left(\a\ot\b X \right)
\eeq where $X$ is independent of
$\a\eqdef\a_{t_1}\ot\a_{t_2}\ot\cdots\ot\a_{t_n}$
and
$\b\eqdef\b_{t_1'}\ot\b_{t_2'}\ot\cdots\ot\b_{t_m'}$.

In the HPO formalism, the decoherence functional
$d$ has been constructed for the special case of
continuous-time projection operators corresponding
to coherent states \cite{IL95}, as discussed
above. The decoherence functional $d(\mu,\nu)$ for
two such continuous-time histories is denoted by
\begin{equation}
  d(\mu,\nu) = tr_{{\mathcal{V}}_{cts}\otimes
  {\mathcal{V}}_{cts}}( P_{|\exp \mu(\cdot)\rangle}
  \otimes P_{|\exp (\nu(\cdot)\rangle} X)
  \end{equation}
where
 \begin{equation}
  X:= \langle 0 | \rho_{-\infty}| 0
  \rangle({\mathcal{S}}_{cts}{\mathcal{U}})^{\dagger}
  \otimes({\mathcal{S}}_{cts}{\mathcal{U}}).
\end{equation}
and the two continuous time projectors $P_{|\exp
\mu(\cdot)\rangle}$ and $P_{|\exp
\nu(\cdot)\rangle}$ correspond to the two
continuous-time histories $\mu$ and $\nu$
respectively.\ The appearance of the operator
${\mathcal{S}}_{cts}{\mathcal{U}}$ in the
expression of the decoherence functional is not
restricted to coherent state propositions, but it
arises in the decoherence functional
\cite{xaris99}.

We will now demonstrate certain relations between
${\mathcal{S}}_{cts}{\mathcal{U}}$ and the three
crucial operators $H, V, S$ of the HPO theory, in
order to emphasise that the appearance of such
operators in the description of the dynamics is
not just a matter of the mathematical formulation
used, but it is also a consequence of the physical
interpretation of the theory.

\subsection{The Appearance of  the Action Operator in  the Decoherence Functional}
The operator ${\mathcal{S}}_{{\rm cts}}$ that
appears in the expression for the $d(\mu,\nu)$ was
defined in \cite{1} as an approximation of the
derivative operator in the sense that
\begin{equation}
  {\mathcal{S}}_{cts}|\exp \nu(\cdot)\rangle =
    |\exp(\nu(\cdot) + \dot{\nu}(\cdot))\rangle,   \label{Scts}
\end{equation}
while the dynamics was introduced by the operator
${\mathcal{U}}$, defined in such way that the
notion of time evolution is encoded in the
expression
\begin{equation}
  e^{\langle \lambda, \dot{\lambda} \rangle}
  e^{\frac{i}{\hbar}H[\lambda]} =
  tr_{{\mathcal{V}}_{cts}}({\mathcal{S}}_{cts}
  {\mathcal{U}}{\cal{P}}_{|\exp \lambda(\cdot)\rangle}) \label{U}
\end{equation}

We expect $V$ and $H$ to play a similar role to
that of ${\mathcal{S}}_{cts}$ and ${\mathcal{U}}$
respectively, inside an expression for the
decoherence functional. To demonstrate this we
will use the type of Fock space construction given
in eqs.\ (\ref{prop1}---\ref{prop2}). In
particular, we use the property
\begin{equation}
 \Gamma(A) |\exp\nu(\cdot)\rangle = |\exp(A\nu(\cdot))\rangle
\end{equation} where $A$ is an operator that acts on the elements $ \nu
(\cdot)$ of the base Hilbert space $\mathcal{H}$,
while the operator $\Gamma(A)$, defined by eq.\
(\ref{prop1}), acts on the coherent states
$|\exp\nu(\cdot)\rangle$ of the Fock space
$e^{\mathcal{H}}$.

We notice that ${\mathcal{U}}$ is related to the
unitary time-evolution eq.\ (\ref{U}) in a similar
way to that of the Hamiltonian operator $H$
\begin{equation}
    e^{isH} |\exp\nu(\cdot) \rangle = \Gamma ( e^{is\omega I})
|\exp\nu(\cdot)\rangle = |\exp(e^{is\omega}
\nu(\cdot) )\rangle
    \label{Action_eisH}
\end{equation}
where $I$ is the unit operator. We also notice
that the action of the operator $e^{isH}$ produces
phase changes, as reflected in the right hand side
of eq.\ (\ref{Action_eisH}) (which has been
calculated for the special case of the simple
harmonic oscillator). Furthermore, when the
operator ${\mathcal{S}}_{cts}$ acts on a coherent
state eq.\ (\ref{Scts}), it transforms it to
another coherent state which involves the addition
to the defining function $\nu(\cdot)$ in a way
that involves the time derivative of $\nu$; and it
is noteworthy that the Liouville operator $V$ acts
in a similar way:
\begin{equation}
 e^{isV} |\exp\nu(\cdot) \rangle =
 \Gamma ( e^{isD}) |\exp\nu(\cdot)\rangle =
 |\exp(e^{isD}\nu (\cdot))\rangle
 \end{equation}
 where
 \begin{equation}
  (e^{isD} \nu)(t)  = \nu(t+s)
\end{equation}
where $D:= -i\frac{d}{dt}$. The operator $e^{isD}$
acts on the base Hilbert space, and corresponds to
the operator $e^{isV}$ under the
$\Gamma$-construction\footnote{An important
property of the Fock construction states that when
there exists a unitary operator $e^{i s A }$
acting on $L^{2}({\mathR})$, there exists a
unitary operator $\Gamma(e^{i s A })$ that acts on
the exponential Fock space.} on the Fock space;
that is, it acts on the vector $\nu(t)$ and
transforms it to another one $ \nu(t+s)$, which,
for each time $t$ is translation by the time
interval $s$.

This suggests that we define the operator $
{\mathcal{A}}_{s} := e^{i s S}$, where $S:=
\int^{+\infty} _{-\infty} (p_{t}\dot{x}_{t}-
H_{t})dt$ is the action operator for the simple
harmonic oscillator, which one expects to be
related to the operator
${\mathcal{S}}_{cts}{\mathcal{U}}$. For this
reason, we write the matrix elements of both
operators and compare them.

The general formula for the matrix elements of an
arbitrary operator ${\mathcal{T}}$ with respect to
the coherent states basis in the history space in
\cite{IL95} is
\begin{eqnarray}
 \langle
\exp \mu(\cdot)| {\mathcal{T}} |\exp
\nu(\cdot)\rangle&=& e^{(\langle \mu,
\frac{\delta}{\delta \bar{\lambda}} \rangle +
\langle \frac{\delta}{\delta \lambda}, \nu
\rangle)} \langle \exp \lambda(\cdot)|
{\mathcal{T}} | \exp \lambda(\cdot) \rangle
{\Big{|}}_{\lambda={\bar{\lambda}}=0}
\end{eqnarray}
hence we need only compare the diagonal matrix
elements of the two operators
${\mathcal{S}}_{cts}{\mathcal{U}}$ and ${\mathcal
A}_s$. Thus we have
\begin{equation}
 \langle
\exp(\lambda(\cdot)|{\mathcal{S}}_{cts}{\mathcal{U}}|\exp(\lambda(\cdot)\rangle
= e^{\langle\lambda, \lambda
+\dot{\lambda}\rangle}
e^{\frac{i}{\hbar}H[\lambda]}
\end{equation}
where $H[\lambda]:=
\int^{\infty}_{-\infty}H(\lambda(t))dt $ and
$H(\lambda):=H(\lambda, \lambda )= \langle \lambda
| H | \lambda \rangle / \langle \lambda | \lambda
\rangle$; and
\begin{equation}
  \langle
  \exp\lambda(\cdot)|{\mathcal{A}}_{s}|\exp\lambda(\cdot)\rangle=
e^{\langle \lambda , e^{is ( \omega I + D )}
\lambda \rangle}
\end{equation}
with
\begin{equation}
  (e^{is( \omega I +  D )} \lambda)(t) = e^{is\omega} \lambda (t+s)
\end{equation}

We also write the diagonal matrix elements of the
Liouville, the Hamiltonian and the action
operators
\begin{eqnarray}
  \langle
  \exp\lambda(\cdot)|V|\exp\lambda(\cdot)\rangle &=&
 \langle \lambda , D\lambda \rangle e^{\langle \lambda , \lambda \rangle}               \\
 \\
  \langle
  \exp\lambda(\cdot)|H|\exp\lambda(\cdot)\rangle &=&
  \langle \lambda , \omega I \lambda \rangle e^{\langle \lambda , \lambda \rangle}       \\
  \\
  \langle
  \exp\lambda(\cdot)|S|\exp\lambda(\cdot)\rangle &=&
  \langle \lambda , (\omega I + D)\lambda \rangle e^{\langle \lambda , \lambda \rangle}
\end{eqnarray}

We can also write both of the above operators on
the history space ${\mathcal{F}}(L^{2}({\mathR}))$
using their corresponding operators on the Hilbert
space $L^{2}(\mathR)$. The $\Gamma$ construction
shows that
\begin{eqnarray}
  {\mathcal{S}}_{cts} &=& \Gamma(1 + i D)  \\
  {\mathcal{S}}_{cts}{\mathcal{U}} &=& \Gamma(1 + i\sigma) \\
  {\mathcal{A}}_{s} &=&\Gamma(e^{is\sigma}) = e^{isd\Gamma(\sigma)} \\
  V &=& d\Gamma(D)              \\
  S &=& d\Gamma(\omega I + D)   \\
  H &=& d\Gamma(I)
\end{eqnarray}
where $\sigma=\omega I + D$, and $I$ is the unit
operator. As expressions of the same function
$\sigma $, the operators ${\mathcal{S}}_{{\rm
cts}}{\mathcal{U}}$ and $ {\mathcal{A}}_{s}$
commute. However, we cannot readily compute their
common spectrum because the operator
${\mathcal{S}}_{{\rm cts}}{\mathcal{U}}$ is not
self-adjoint.

We might speculate that the value of the
decoherence functional is maximised for a
continuous-time projector that corresponds to a
coarse graining around the classical path. Indeed,
if we take such a generic projection operator $P$,
we expect that it should commute with the operator
${\mathcal{S}}_{cts}{\mathcal{U}}$. In this
context, we noticed earlier that the projection
operator which corresponds to a classical solution
$(f,h)$ commutes with the action operator
\begin{equation}
  [{\mathcal{S}}_{cts}{\mathcal{U}} , P_{(f,h)}] = 0.
\end{equation}
Finally, this argument should be compared with the
similar condition for classical histories:
\begin{equation}
  \{ S_{h} , F_{C} \} (\gamma_{cl}) = 0.
\end{equation}

\chapter{A Study of a Free Relativistic Quantum Field}

\section{Introduction}

We wish now to extend the discussion to the HPO theory of a free
scalar field. Hartle \cite{Har93a} proposed a consistent histories
approach to quantum field theory based on path integrals, and
Blencowe \cite{Ble91} gave a careful analysis of the use of class
operators.  However, almost nothing has been said about the HPO
scheme in this context, and we shall now briefly present the
necessary developments.  The resemblance of the history version of
quantum mechanics (`field theory in zero spatial dimensions') to a
canonical field theory in one spatial dimension suggests that the
history version of quantum field theory in three spatial dimensions
should resemble canonical quantum field theory in {\em four\/}
spatial dimensions.  We shall see that this expectation is fully
justified.

\section{The Canonical History Algebra}

The first step in constructing an HPO version of quantum field
theory is to foliate four-dimensional Minkowski space-time with the
aid of a time-like vector $n^\mu$ that is normalised by
$\eta_{\mu\nu}n^\mu n^\nu=1$, where the signature of the Minkowski
metric $\eta_{\mu\nu}$ has been chosen as $(+,-,-,-)$. The canonical
commutation relations for a standard bosonic quantum field theory
(the analogue of Eq.\ (\ref{CCR})) in three spatial dimensions are
\begin{eqnarray}
    &&{[\,}\phi(\underline{x}_1),\,\phi(\underline{x}_2)\,]=0
                                                    \label{[phiphi]} \\
    &&{[\,}\pi(\underline{x}_1),\,\pi(\underline{x}_2)\,]=0
                                                    \label{[pipi]} \\
    &&{[\,}\phi(\underline{x}_1),\,\pi(\underline{x}_2)\,]=
i\hbar\delta^3(\underline{x}_1-\underline{x}_2). \label{[phipi]}
\end{eqnarray}
 In
constructing the associated HPO theory we shall assume that the
passage from the canonical algebra Eq.\ (\ref{CCR}) to the history
algebra Eqs.\ (\ref{ctsHWxx})--(\ref{ctsHWxp}) is reflected in the
field theory case by passing from Eqs.\
(\ref{[phiphi]})--(\ref{[phipi]}) to
\begin{eqnarray}
&&{[\,}\phi_{t_1}(\underline{x}_1),\,\phi_{t_2}(\underline{x}_2)\,]=0
                                        \label{[tphiphi]}       \\
    &&{[\,}\pi_{t_1}(\underline{x}_1),\,\pi_{t_2}(\underline{x}_2)\,]=0
                                        \label{[tpipi]}         \\
    &&{[\,}\phi_{t_1}(\underline{x}_1),\,\pi_{t_2}(\underline{x}_2)\,]=
        i\hbar\delta(t_1-t_2)\delta^3(\underline{x}_1-\underline{x}_2)
                                        \label{[tphipi]}
\end{eqnarray}
where, for each $t\in\mathR$, the fields $\phi_t(\underline{x})$ and
$\pi_t(\underline{x})$ are associated with the spacelike
hypersurface $(n,t)$ whose normal vector is $n$ and whose foliation
parameter is $t$; in particular, the three-vector $\underline{x}$ in
$\phi_t(\underline{x})$ or $\pi_t(\underline{x})$ denotes a vector
in this space.

    In using this algebra, we have in mind a representation that is
some type of continuous tensor product $\otimes_{t\in\mathR}{\cal
H}_t$ where each ${\cal H}_t$ carries a representation of the
standard canonical commutation relations Eqs.\
(\ref{[phiphi]})--(\ref{[phipi]}) for a scalar field theory
associated with the given spacetime foliation. However, to emphasise
the underlying spacetime picture it is convenient to rewrite Eqs.\
(\ref{[tphiphi]})--(\ref{[tphipi]}) in terms of four-vectors $X$ and
$Y$ as
\begin{eqnarray}
    &&{[\,}\phi(X),\,\phi(Y)\,]=0               \label{[phiXphiY]}\\
    &&{[\,}\pi(X),\,\pi(Y)\,]=0                 \label{[piXpiY]}  \\
    &&{[\,}\phi(X),\,\pi(Y)\,]=i\hbar\delta^4(X-Y).\label{[phiXpiY]}
\end{eqnarray}
In relating these expressions to those in Eqs.\
(\ref{[tphiphi]})--(\ref{[tphipi]}) the three-vector $\underline{x}$
may be equated with a four-vector $x_n$ that satisfies $n\cdot
x_n=0$ (the dot product is taken with respect to the Minkowski
metric $\eta_{\mu\nu}$) so that the pair
$(t,\underline{x})\in\mathR\times\mathR^3$ is associated with the
spacetime point $X=tn+x_n$ (in particular, $t=n\cdot X$). Note,
however, that the covariant-looking nature of these expressions is
deceptive and it is not correct to assume {\em a priori\/} that the
fields $\phi(X)$ and $\pi(Y)$ transform as spacetime scalars under
the action of some `external' spacetime Poincar\'e group that acts
on the $X$ and $Y$ labels---as things stand there is an implicit $n$
label on both $\phi$ and $\pi$. We shall return to this question
later.

\section{The Hamiltonian Algebra}
The key idea of our HPO approach to quantum field theory is that the
physically-relevant representation of the canonical history algebra
Eqs.\ (\ref{[tphiphi]})--(\ref{[tphipi]}) [or, equivalently, Eqs.\
(\ref{[phiXphiY]})--(\ref{[phiXpiY]})] is to be selected by
requiring the existence of operators that represent history
propositions about temporal averages of the energy defined with
respect to the chosen spacetime foliation. Thus, for a fixed
foliation vector $n$, we seek a family of `internal' Hamiltonians
$H_{n,t}$, $t\in\mathR$, whose explicit formal form ({\em i.e.}, the
analogue of Eq.\ (\ref{Def:Ht})) can be deduced from the standard
quantum field theory expression to be
\begin{equation}
    H_{n,t}:={1\over 2}\int d^4X\left\{\pi(X)^2+
    (n^\mu n^\nu-\eta^{\mu\nu})\partial_\mu\phi(X)\partial_\nu\phi(X)
        +m^2\phi(X)^2\right\}\delta(t-n\cdot X).\label{Def:Hnt}
\end{equation}
The analogous, temporally-averaged object is
\begin{eqnarray}
    H_n(\chi)&:=&\int_{-\infty}^\infty dt\,\chi(t) H_{n,t}
                                                \label{Def:Hnchi}\\
        &=&{1\over 2}\int d^4X\left\{\pi(X)^2+
    (n^\mu n^\nu-\eta^{\mu\nu})\partial_\mu\phi(X)\partial_\nu\phi(X)
        +m^2\phi(X)^2\right\}\chi(n\cdot X)             \nonumber
\end{eqnarray}
where $\chi$ is a real-valued test function.

    As in the discussion above of the simple harmonic oscillator,
the next step is to consider the commutator algebra that would be
satisfied by the operators $H_n(\chi)$ {\em if\/} they existed.
These field-theoretic analogues of Eqs.\
(\ref{[Hchixt]})--(\ref{[Hchi1Hchi2]}) are readily computed as
\begin{eqnarray}
    &&{[\,}H_n(\chi),\,\phi(X)\,]=-i\hbar\chi(n\cdot X)\pi(X)
                                                    \label{[HchiphiX]}\\
    &&{[\,}H_n(\chi),\,\pi(X)\,]=i\hbar\chi(n\cdot X) K_n\phi(X)
                                                    \label{[HchipiX]}\\
    &&{[\,}H_n(\chi_1),\,H_n(\chi_2)\,]=0
\end{eqnarray}
where $K_n$ denotes the partial differential operator
\begin{equation}
(K_nf)(X):=\left[(\eta^{\mu\nu}-n^\mu n^\nu)
\partial_\mu\partial_\nu +m^2 \right]f(X). \label{Def:Kn}
\end{equation}

    The exponentiated form of Eqs.\
(\ref{[HchiphiX]})--(\ref{[HchipiX]}) is
\begin{eqnarray}
 &e&^{iH_n(\chi)/\hbar}\,\phi(X)\,e^{-iH_n(\chi)/\hbar}= \nonumber\\
&=& \cos\left[\chi(n\cdot X)\sqrt{K_n}\right]\phi(X)
        +{1\over\sqrt K_n}\sin\left[\chi(n\cdot X)
        \sqrt{K_n}\right]\pi(X) \hspace{2cm}                \label{AutphiX} \\
 \\
 &e& ^{iH_n(\chi)/\hbar}\,\pi(X)\,e^{-iH_n(\chi)/\hbar}= \nonumber\\
 &=& -\sqrt{K_n}\sin\left[\chi(n\cdot X)\sqrt{K_n}\right]\phi(X)
+\cos\left[\chi(n\cdot X)\sqrt{K_n}\right]\pi(X)    \label{AutpiX}
\end{eqnarray}
where the square-root operator $\sqrt{K_n}$, and functions thereof,
can be defined rigorously using the spectral theory of the
self-adjoint, partial differential operator $K_n$ on the Hilbert
space $L^2(\mathR^4,d^4X)$. Note that the expression $\chi(n\cdot
X)\sqrt{K_n}$ is unambiguous since, viewed as an operator on
$L^2(\mathR^4,d^4X)$, multiplication by $\chi(n\cdot X)$ commutes
with $K_n$.

\section{The Fock Space Representation}
The right hand side of Eqs.\ (\ref{AutphiX})--(\ref{AutpiX}) defines
an automorphism of the CHA Eqs.\
(\ref{[phiXphiY]})--(\ref{[phiXpiY]}) and the task is to find a
representation of the latter in which these automorphisms are
unitarily implemented. To this end, define new operators
\begin{eqnarray}
q(X)&:=&K^{1/4}_n\phi(X) \label{Def:qX}\\ p(X)&:=&K^{-1/4}_n\pi(X)
\label{Def:pX}
\end{eqnarray} and
\begin{equation}
    b(X):={1\over\sqrt2}\Big(q(X)+ip(X)\Big)=
      {1\over\sqrt2}\Big(K^{1/4}_n\phi(X)+iK^{-1/4}_n\pi(X)\Big)
                                                \label{Def:bX}
\end{equation}
which satisfy
\begin{eqnarray}
        &&{[\,}b(X),\,b(Y)\,]=0                 \\
        &&{[\,}b^\dagger(X),\,b^\dagger(Y)\,]=0     \\
        &&{[\,}b(X),\,b^\dagger(Y)]=\hbar\delta^4(X-Y).
\end{eqnarray}

Then
\begin{eqnarray}
 &e& ^{iH_n(\chi)/\hbar}\,q(X)\,e^{-iH_n(\chi)/\hbar}= \nonumber\\
 &=& \cos\left[\chi(n\cdot X)\sqrt{K_n}\right]q(X)
            + \sin\left[\chi(n\cdot X)\sqrt{K_n}\right]p(X)\\
 \\
 &e& ^{iH_n(\chi)/\hbar}\,p(X)\,e^{-iH_n(\chi)/\hbar}= \nonumber\\
 &=& -\sin\left[\chi(n\cdot X)\sqrt{K_n}\right]q(X)
            +\cos\left[\chi(n\cdot X)\sqrt{K_n}\right]p(X)
\end{eqnarray}
and so
\begin{equation}
    e^{iH_n(\chi)/\hbar}\,b(X)\,e^{-iH_n(\chi)/\hbar}=
        e^{-i\chi(n\cdot X)\sqrt{K_n}}\,b(X).   \label{Autb(X)}
\end{equation}
However, the operator defined on $L^2(\mathR^4)$ by
\begin{equation}
    (U(\chi)\psi)(X):=e^{-i\chi(n\cdot X)\sqrt{K_n}}\psi(X)
\end{equation}
is unitary, and hence---using the same type of argument invoked
earlier for the simple harmonic oscillator---we conclude that the
desired quantities $H_n(\chi)$ exist as self-adjoint operators on
the Fock space ${\cal F}[L^2(\mathR^4,d^4X)]$ associated with the
creation and annihilation operators $b^\dagger(X)$ and $b(X)$. The
spectral projectors of these operators then represent propositions
about the time-averaged value of the energy in the spacetime
foliation determined by $n$.

\section{The Question of External Lorentz Invariance}
An important part of standard quantum field theory is a proof of
invariance under the Poincar\'e group---something that, in the
canonical formalism, is not totally trivial since the
Schr\"odinger-picture fields depend on the reference frame ({\em
i.e.}, the spacetime foliation). The key ingredient is a
construction of the generators of the Poincar\'e group as explicit
functions of the canonical field variables; in practice, the first
step is often to construct the Heisenberg-picture fields with the
aid of the Hamiltonian, and then to demonstrate manifest Poincar\'e
covariance within that framework.  The canonical fields associated
with any spacelike surface in a particular Lorentz frame can then be
obtained by restricting the Heisenberg fields (and their normal
derivatives) to the surface.

    When considering the role of the Poincar\'e group in the HPO
picture of consistent histories, the starting point is the
observation that, heuristically speaking, for a given foliation
vector $n$---and for each value of the associated time $t$---there
will be a Hilbert space ${\cal H}_t$ carrying an independent copy of
the standard quantum field theory. In particular, therefore, for
fixed $n$, there will be a representation of the Poincar\'e group
associated with each spacelike slice $(n,t)$, $t\in\mathR$.  Thus if
$A_a$, $a=1,2,\ldots,10$ denote the generators of the Poincar\'e
group, there should exist a family of operators $A_t^a$ which, for
each $t\in\mathR$, generate the `{\em internal\/}' Poincar\'e group
${\cal P}_{n,t}$ associated with the slice $(n,t)$.  These operators
will satisfy a `temporally gauged' version of the Poincar\'e
algebra.  More precisely, if $C^{ab}{}_c$ are the structure
constants of the Poincar\'e group, so that
\begin{equation}
    [\,A^a,\,A^b\,]=iC^{ab}{}_cA^c,
\end{equation}
then the algebra satisfied by the history theory operators $A_t^a$
is
\begin{equation}
    [\,A^a_t,\,A^b_s\,]=i\delta(t-s) C^{ab}{}_cA_t^c
\end{equation}
which, of course, reflects the way in which the canonical
commutation relations Eqs.\ (\ref{[phiphi]}--\ref{[phipi]}) are
replaced by Eqs.\ (\ref{[tphiphi]}--\ref{[tphipi]}) in the history
theory.

    As always in quantum theory, the energy operator is of
particular importance, and in the present case we have a family of
Hamiltonian operators $H_{n,t}$, $t\in\mathR$, which are related to
the generators $P_{n,t}^\mu$ of translations for the quantum field
theory associated with the hypersurface $(n,t)$ by
\begin{equation}
    H_{n,t}=n_\mu P_{n,t}^\mu.
\end{equation}
In fact, it is straightforward to show that
\begin{equation}
    P_{n,t}^\mu=n^\mu H_{n,t}+\int d^4X\,\delta(t-n\cdot X)
        (n^\mu n\cdot\partial\phi-\partial^\mu\phi)\pi
\end{equation}
which suggests that, as would be expected, the components of
$P_{n,t}^\mu$ normal to $n$ act are the generators of spatial
translations in the hypersurface $(n,t)$. Indeed, Eq.\
(\ref{[HchiphiX]}) generalises to
\begin{equation}
    {[\,}P_n^\mu(\chi),\,\phi(X)\,]=-i\hbar\chi(n\cdot X)
    \big\{n_\mu\pi(X)+(\partial_\mu\phi(X)-
        n_\mu\, n\cdot\partial\phi(X))\big\}.
\end{equation}
Similarly, the `temporally gauged' Lorentz generators satisfy
\begin{eqnarray}
\lefteqn{{[\,}J_{n,t}^{\mu\nu},\,\phi(X)\,] = }         \\
        & & \qquad i\hbar\delta(t-n\cdot X)
        \big\{X^\mu(\partial^\nu\phi-n^\nu n\cdot\partial\phi)
            - X^\nu(\partial^\mu\phi-n^\mu n\cdot\partial\phi)
            -(X^\mu n^\nu-X^\nu n^\mu)\pi\big\}\nonumber.
\end{eqnarray}

    As emphasised above, each generator of the group ${\cal
P}_{n,t}$ acts `internally' in the Hilbert space ${\cal H}_t$; in
particular, this is true of the Hamiltonian, which (modulo the need
to smear in $t$) generates translations along an `internal' time
label $s$ that is to be associated with each leaf $(n,t)$ of the
foliation. It is important to note that $H_{n,t}$ does {\em not\/}
generate translations along the `exernal' time parameter $t$ that
appears in the CHA Eqs.\ (\ref{[tphiphi]}--\ref{[tphipi]}) and which
labels the spacelike surface (of course, there is an analogous
statement for the Hamiltonians $H_t$ in the HPO model of the simple
harmonic oscillator considered earlier). The existence of these
internal Poincar\'e groups is sufficient to guarantee covariance of
physical quantities, such as transition amplitudes, that can be
calculated in the class operator version of the theory.

    However, the HPO formalism admits an additional type of
Poincar\'e group---what we shall call the `{\em external\/}'
Poincar\'e group---which is defined to act on the pair of labels
$(\underline{x},t)$ that appear in the CHA Eqs.\
(\ref{[tphiphi]}--\ref{[tphipi]}). Thus these labels include the
`external' time parameter $t$ that specifies the leaf $(n,t)$ of the
foliation associated with the timelike vector $n$. In the context of
the covariant-looking version Eqs.\
(\ref{[phiXphiY]}--\ref{[phiXpiY]}) of the CHA, the main question is
whether the fields $\phi(X)$ and $\pi(X)$ transform in a covariant
way under this external group.

    As far as the field $\phi(X)$ is concerned it seems reasonable
to consider the possibility that this may an external scalar in the
sense that there exists a unitary representation $U(\Lambda)$ of the
external Lorentz group $U(\Lambda)$ such that
\begin{equation}
    U(\Lambda)\phi(X)U(\Lambda)^{-1}=\phi(\Lambda X).
\end{equation}
The spectral projectors of the (suitably smeared) operators
$\phi(X)$ then represent propositions about the values of the
spacetime field in a covariant way.

    However, the situation for the
field momentum $\pi(X)$ is different since this is intrinsically
associated with the timelike vector $n$.  Indeed, the natural thing
would be to require the existence of a {\em family\/} of operators
$\pi_n(X)$ where $n$ lies in the hyperboloid of all timelike
(future-pointing) vectors, and such that
\begin{equation}
        U(\Lambda)\pi_n(X)U(\Lambda)^{-1}=
                    \pi_{\Lambda n}(\Lambda X).
\end{equation}
The next step in demonstrating external Poincar\'e covariance would
be to extend the algebra (\ref{[phiXphiY]}--\ref{[phiXpiY]}) to
include the $n$ parameter on the $\pi$ field; in particular, one
would need to specify the commutator $[\,\pi_n(X),\,\pi_m(Y)\,]$,
but it is not obvious {\em a priori} what this should be.

    Another possibility would be to try to combine the Heisenberg
picture---and its associated `internal' time $s$---with the external
time parameter $t$ of the spacetime foliation to give some scheme
that was manifestly covariant in the context of a five-dimensional
space with signature $(+++,--)$ associated with the variables
$(\underline{x},t,s)$. However, we do not know if this is possible and
the demonstration of external Poincar\'e covariance, if it exists,
remains the subject for future research.

\chapter{Conclusions}
    We have discussed the introduction of continuous-time histories
within the `HPO' version of the consistent-histories formalism in
which propositions about histories of the system are represented by
projection operators on a `history' Hilbert space. The history
algebra (whose representations specify this space) for a particle
moving in one dimension is isomorphic to the canonical commutation
relations for a one-dimensional quantum field theory, thus
allowing the history theory to be studied using techniques drawn from
quantum field theory. In particular, we have shown how the problem
of the existence of infinitely many inequivalent representations of
the history algebra can be solved by requiring the existence of
operators whose spectral projectors represent propositions about
time-averages of the energy.

    We have examined the example of the simple harmonic oscillator,
in one dimension, within the History Projection Operator
formulation of the consistent-histories scheme. We defined the
action operator as the quantum analogue of the classical
Hamilton action functional and we have proved its existence by
finding a representation on the
${\mathcal{F}}({\mathcal{L}}^{2}({\mathbf{R}}))$ space of the
history algebra. We have shown that the action operator is the
generator of two types of time transformations: translations in
time from one Hilbert space ${\mathcal{H}}_{t}$, labeled by the
time parameter $t$, to another Hilbert space with a different
label $t$, and phase changes in time with respect to the time
parameter $s$ of the standard Heisenberg-time evolution that
acts in each individual Hilbert space ${\mathcal{H}}_{t}$. We
have expressed the action operator in terms of the Liouville and
Hamiltonian operators---which are the generators of the two
types of time transformation---and which correspond to the
kinematics and the dynamics of the theory respectively.

    We have constructed continuous-time classical histories defined
on the continuous Cartesian product of copies of the phase space
and demonstrated an analogous expression to the classical
Hamilton's equations.

    We have shown that the action operator commutes with the
defining operator of the decoherence functional, thus appearing
in the expression for the dynamics of the theory, as would have
been expected.

	Finally, we have shown how the HPO scheme can be extended to
the history version of canonical quantum field theory. We discussed
the difference between the `internal' and `external' Poincar\'e
groups and indicated how the former are implemented in the formalism.
A major challenge for future research is to construct an HPO quantum
field theory which is manifestly covariant under this external
symmetry group.

One of the major reasons for undertaking this study was to
provide new tools for tackling the recalcitrant problem of
constructing a manifestly covariant quantum field theory in the
consistent histories formalism. Work on this problem is now in
progress with the expectation that the Hamiltonian and Liouville
operators will play a central role in the proof of explicit
Poincar\'e invariance of the theory.

 \end{document}